\documentclass[%
 reprint,
 amsmath,amssymb,
 aps,
longbibliography
]{revtex4-1}

\usepackage{graphicx}
\graphicspath{{./images/}}
\usepackage{dcolumn}
\usepackage{bm}
\usepackage{enumitem}

\begin{document}

\preprint{AIP/123-QED}

\title[The stability of slowly evaporating thin liquid films of binary mixtures]{The stability of slowly evaporating thin liquid films of binary mixtures}

\author{R. K. Nazareth}
 \altaffiliation[]{Institute for Multiscale Thermofluids, School of Engineering, The University of Edinburgh, UK.} 

\author{G. Karapetsas}%
 \altaffiliation[]{Department of Chemical Engineering, Aristotle University of Thessaloniki, Greece.}
 
\author{K. Sefiane}
 \altaffiliation[]{Institute for Multiscale Thermofluids, School of Engineering, The University of Edinburgh, UK.}

\author{O. Matar}
 \altaffiliation[]{Department of Chemical Engineering, Imperial College London, UK.}

\author{P. Valluri}
 \altaffiliation[]{Institute for Multiscale Thermofluids, School of Engineering, The University of Edinburgh, UK.}

\date{\today}

\begin{abstract}
We consider the evaporation of a thin liquid layer which consists of a binary mixture of volatile liquids. The mixture is on top of a heated substrate and in contact with the gas phase that consists of the same vapour as the binary mixture. The effect of thermocapillarity, solutocapillarity and the van der Waals interactions are considered. We derive the long-wave evolution equations for the free interface and the volume fraction that govern the two-dimensional stability of the layer subject to the above coupled mechanisms and perform a linear stability analysis. Our results demonstrate two modes of instabilities, a monotonic instability mode and an oscillatory instability mode. We supplement our results from stability analysis with transient simulations to examine the dynamics in the nonlinear regime and analyse how these instabilities evolve with time. More precisely we discuss how the effect of relative volatility along with the competition between thermal and solutal Marangoni effect defines the mode of instability that develops during the evaporation of the liquid layer due to preferential evaporation of one of the components.
\end{abstract}

\pacs{Valid PACS appear here}
\maketitle

\section{Introduction}

The dynamics of binary films subjected to temperature and solute concentration gradients is an important problem which has widespread technological applications like coating, wetting and cooling processes. The stability of thin liquid films was reviewed by Oron \textit{et al.} \cite{Oron1997} and Craster \& Matar \cite{Crester2009}.

Pearson \cite{Pearson1958} was the first to address instabilities in liquid layers driven by surface tension gradients. He observed that drying paint films often display steady cellular circulatory flow similar to those examined by B\'enard \cite{Benard1901} in liquid layers heated from below. In the former case, the cellular pattern was observed even when the free surface was on the underside of the paint layer and the gravity vector was effectively inverted. Therefore, Pearson \cite{Pearson1958} concluded that density gradient, as proposed by Rayleigh \cite{Rayleigh1916} to explain B\'ernard hexagonal cellular pattern, cannot be the mechanism causing the instabilities in this case and proposed surface tension forces as the driving force of the cellular patterns. Pearson \cite{Pearson1958} performed a stability analysis on a liquid layer heated from below by means of small-disturbance theory, similar to that developed by Rayleigh \cite{Rayleigh1916}. In his analysis surface tension was assumed a linearly decreasing function of temperature, the interface was non-deformable and gravity was neglected. Pearson \cite{Pearson1958} derived critical values of the Marangoni number corresponding to the case of convective instabilities. Pearson's \cite{Pearson1958} stability analysis was extended by Scriven \& Sternling \cite{Sternling1964} by accounting for the possibility of shape deformations of the free surface. They found that there is no critical Marangoni number for the onset of stationary instability and that the limiting case of ``zero wave-number'' (i.e. waves of very large wavelengths as in a thin film) is always unstable. Scriven \& Sternling \cite{Sternling1964} also provided a criterion to distinguish visually whether buoyancy or surface tension dominate cellular convection in liquid pools. While in surface tension driven instabilities, the flow is towards the free surface in shallow sections and away in deeper sections, this relationship is just the opposite in buoyancy-driven flows, as observed by Jeffreys \cite{Jeffreys1951}. 

William \& Davis \cite{Williams1982} posed a nonlinear stability theory based on the long-wave nature of the response. They derived a partial differential equation which describes the evolution of the interface shape subject to surface tension, viscous forces, and the van der Waals attractions. They found that the nonlinear measure of the rupture time is always smaller than the equivalent measure given by a linearized theory. Burelbach \textit{et al.} \cite{Burelbach1988} extended the nonlinear theory developed by William \& Davis \cite{Williams1982} to include evaporative, thermocapillary, and non-equilibrium effects, in addition to disjoining pressures induced by van der Waals attractions. They derived long-wave evolution equations for the interface shapes that govern the stability of the layers subject to the above coupled mechanisms to investigate film instabilities and rupture. They show that increasing the degree of thermocapillarity decreases the time for rupture of the film. Goussis \& Kelly \cite{Goussis1990} analysed the importance of the layer thickness on thermocapillary instabilities. In sufficiently thick layers, instabilities can take the form of relatively short wavelengths which are of the order of the layer's depth, as Pearson \cite{Pearson1958} demonstrated. This instability is associated with the interaction of the basic temperature with the perturbed velocity field and effects of convection are important. For sufficiently thin films, surface tension stabilizes short wavelengths so the instability takes the form of large wavelengths disturbances. This instability is associated with the modification of the basic temperature by the deformation of the free surface. 

Stability of evaporating films due to solutal effects was also considered by many authors. Hatziavramidis \cite{Hatziavramidis1992} performed linear stability analysis on evaporating films with soluble surfactant considering the flow effects arising from surface tension gradients due to temperature and concentration variations, in addition to van der Waals forces and surface tension. They quantified the effect of surfactant in terms of its adsorption at the liquid-gas interface. They found that flows driven by surface tension gradients originating from surface concentration variations are in a direction opposite to similar flows originating from surface temperature variations. The former usually dominate; they are destabilizing for condensing films and stabilizing for evaporating films. Danov et al. \cite{Danov1998} investigated the dynamics of an evaporating film in the presence of dissolved surfactant using lubrication approximation taking into account interfacial mass loss due to evaporation, the van der Waals attraction, the Marangoni effect due to thermal and concentration variations, and the effect of interfacial viscosity on film stability. They found that increasing the initial surfactant concentration stabilises the film only up to the moment of reaching tangential immobility of the interface due to the increase of its interfacial viscosity and elasticity. After that, the additional increase of surfactant concentration leads only to a decrease of interfacial tension, lowering the film stability. 

Lin et al. \cite{LinUen2000} investigated the effects of soluble surfactant on the dynamic rupture of thim liquid films. They adopted a generalized Frumkin model to simulate the adsorption/desorption kinetics of the soluble surfactant between the surface and the bulk phases. They show by means of numerical simulations that the liquid film system with soluble surfactant is more unstable than that with insoluble surfactant. They found that surfactant solubility increases as absorption/desorption rate, activation energy, and bulk diffusion increase, which causes the film system to becomes unstable, and the surfactant solubility decreases as the rate of equilibrium and interaction among molecules increase, which therefore stabilizes the film. They found that an increase of relative surface concentration initially result in a decrease of corresponding shear drag forces which enhance the Marangoni effect and a further increase of relative surface concentration result in an increase of the corresponding shear drag force which weaken the Marangoni effect and result in a reduction of the interfacial stability. Yiantsios \& Higgins \cite{YiantsiosHiggins2010} analysed a mechanism of Marangoni instability in evaporating films with soluble surfactant. Using linear stability analysis they show that the instability will manifest itself provided that an appropriate Marangoni number is relatively large and the surfactant solubility in the bulk is large as well. They found that low solubility in the bulk, diffusion, and the effect of surfactant on interfacial mobility through the surface viscosity suppress disturbance growth. They confirm the results using direct numerical simulations of the nonlinear evolution equations. 

Mikishev \& Nepomnyashchy \cite{MikishevNepomnyashchy2013, MikishevNepomnyashchy2014} studied the stability of an evaporating film with insoluble surfactant distributed over the free deformable interface. The insoluble surfactant hinders the evaporation, and mass flux through the interface are a decreasing function of surfactant concentration. Using a one-sided model and the long-wave approximation under the assumption of a slow time evolution, linear stability analysis of the base state is performed for long-wave disturbances using frozen interface approximation. The authors analyse the cases of quasi-equilibrium and non-equilibrium evaporation and found monotonic and oscillatory instability modes. Instability thresholds were determined and critical Marangoni numbers were found for monotonic and oscillatory instabilities using the one-sided model and linear stability analysis for different values of kinetic resistance parameter.

Overdiep \cite{Overdiep1986} developed integro-differential equations and performed experiments to study the levelling process in paint films. They found that the solutal Marangoni effect drive the liquid from the trough with higher concentration of resin to the crest, levelling the perturbation. Howison et al. \cite{Howison1997} developed a mathematical model based on classical lubrication theory for a drying paint layer consisting of a non-volatile resin and a volatile solvent. They considered the effects of variable surface tension, viscosity, solvent diffusivity and solvent evaporation rate. They provide an analytical description of the `reversal' of an initial perturbation to the thickness of the layer and the appearance of a perturbation to an initially flat layer caused by an initial perturbation to the concentration of solvent. Eres et al. \cite{Eres1999} presented a three-dimensional mathematical and numerical model based on the lubrication approximation for the flow of drying paint films on horizontal substrates. They consider the effects of surface tension and gravitational forces as well as surface tension gradient effects which arise due to solvent evaporation and the dependence of viscosity, diffusivity, and evaporation rate on resin concentration. Their model demonstrates the effect of surface tension gradients due to compositional changes in a three-dimensional flow field.

Most of the work on binary films has been focused on the Soret effect. Takashima \cite{Takashima1979} examined the onset of instability in a horizontal binary film subjected to a vertical temperature gradient taking account the Soret effect using linear stability theory. Takashima \cite{Takashima1980} extended his previous work to include the possibility of overstability (oscillatory instability). Joo \cite{Joo1995} analysed the stability of a binary film heated from above. The heat transfer is driven by the vertical temperature gradient. The mass flux is induced by the Soret effect. The instability is driven by solutocapillarity and retarded by thermocapillarity. Small-wavenumber and the Pearson-type instabilities are studied. Oscillatory instability can exist when the thermocapillarity is destabilizing and the solutocapillarity is stabilizing. Podolny et al. \cite{PodolnyNepomnyashchy2005} investigated the long-wave Marangoni instability in a binary film in the limit of small Biot number. The surface deformation and the Soret effect are both taken into account. They characterized the problem by two distinct asymptotic limits for the disturbance wavenumber using the Biot number, which are caused by the action of two instability mechanisms, the thermocapillary and solutocapillary effects. They found a new oscillatory mode for sufficiently small values of the Galileo number.  Podolny et al. \cite{PodolnyNepomnyashchy2006} investigated the long-wave Marangoni instability in binary film in the presence of the Soret effect in the case of finite Biot numbers. Long-wave monotonic and oscillatory instability modes are found in various parameter domains using linear stability analysis. Stable supercritical pattenrs are investigated in the limit of low gravity using weakly nonlinear analysis. Supercritical standing and travelling waves are noted. Borcia et al. \cite{BorciaBestehorn2006} examined long-wave instabilities in binary films accounting to the Soret effect. Linear stability analysis reveals monotonic and oscillatory instabilities. Typical structures such as static or soliton like traveling drops are analysed using 3D non-linear simulations. Zhang et al. \cite{ZhangOron2007} examined Marangoni instabilities in binary films in the presence of the Soret effect and evaporation using NaCl/water mixtures. They investigated the flow pattern formation using a shadow-graph method for a set of substrate temperatures and solute concentrations in non-deformable interface. They found patterns mainly composed of polygons and rolls. They found that evaporation affects the pattern formation mainly at early stages and the Soret effect becomes important at later stages. The strength of convection increases with the initial solute concentration and the substrate temperature. Machrafi et al. \cite{MachrafiDauby2010} performed linear stability analysis on a horizontal binary film using water/ethanol mixtures with the evaporation of water being neglected. They calculated neutral (monotonic) stability curves in terms of solutal/thermal Marangoni/Rayleigh numbers as a function of the wavenumber for different values of the ratio of the gas and liquid layer thicknesses. For a 10 wt.\% water-ethanol mixture they found the solutal Marangoni effect as the most important instability mechanism. Bestehorn \& Borcia \cite{BestehornBorcia2010} studied film instabilities in binary films with deformable interface and an externally applied vertical temperature gradient using lubrication theory. Using linear stability analysis they showed that the monotonic long-wave instability may turn into an oscillatory one if the two components have a different surface tension and if the Soret coefficient establishes a stabilizing vertical concentration gradient. They also discussed a real system consisting of a water/isopropanol mixture. 

This work presents an analytical model to investigate the stability and dynamics of the evaporation of an horizontal thin liquid layer composed of a binary mixture of volatile liquids heated from below. The long-wave approximation is used to derive the evolution equations for the free interface and the concentration of the components that govern the two-dimensional stability of the layer. The effect of evaporation of both components, thermo- and solutocapillarity and the van der Waals attraction are considered. Crucially, we examine the effect of the relative volatility of the components of the binary mixture and relevant flow maps are produced. A linear stability analysis is performed to derive the growth rate of the instabilities for the case of quasi-equilibrium evaporation and non-equilibrium evaporation. The developed linear theory describes two modes of instabilities, a monotonic instability mode and an oscillatory instability mode. Further, by means of transient simulations the dependence of these instabilities on the destabilising effects considered is analysed. The transient simulations also help investigate the dynamics in the non-linear regime.

\section{Problem formulation}

The evaporation of a thin liquid layer which consists of a mixture of volatile liquids A and B is investigated. The volatilities of the components are dependent on their respective vapour (saturation) pressures, with the component with highest vapour pressure exhibiting the highest volatility. The mixture is assumed to be ideal while the liquid layer is considered to be Newtonian, with density $\tilde{\rho}$, specific heat capacity $\tilde{c}_p$, thermal conductivity $\tilde{\lambda}$, and viscosity $\tilde{\mu}$, which depend on the local volume fraction of the two volatile components; the tildes stand for dimensional quantities. The surface tension $\tilde{\sigma}$, also depends on the local volume fraction as well as the local temperature given by Eqs. \ref{eq:surface_tension} and \ref{eq:surface_tension_b}. The liquid layer is on the top of a horizontal, uniformly heated solid substrate and is in contact with the gas phase with average bulk temperature $\tilde{T}_g$; the gas consists of the vapour of the binary mixture. It is assumed that, initially, the liquid layer has thickness $\tilde{H}_o$ and length $\tilde{L}_o$. In the present work, it is considered a very thin liquid layer and therefore $\tilde{L}_o$ greatly exceeds $\tilde{H}_o$ so that the ratio, $\epsilon = \tilde{H}_o/\tilde{L}_o$, is assumed to be very small. The latter assumption permits the use of lubrication theory, which will be employed below to derive a set of evolution equations that govern the evaporation process.

\begin{figure}
	\centerline{\includegraphics[width=8cm]{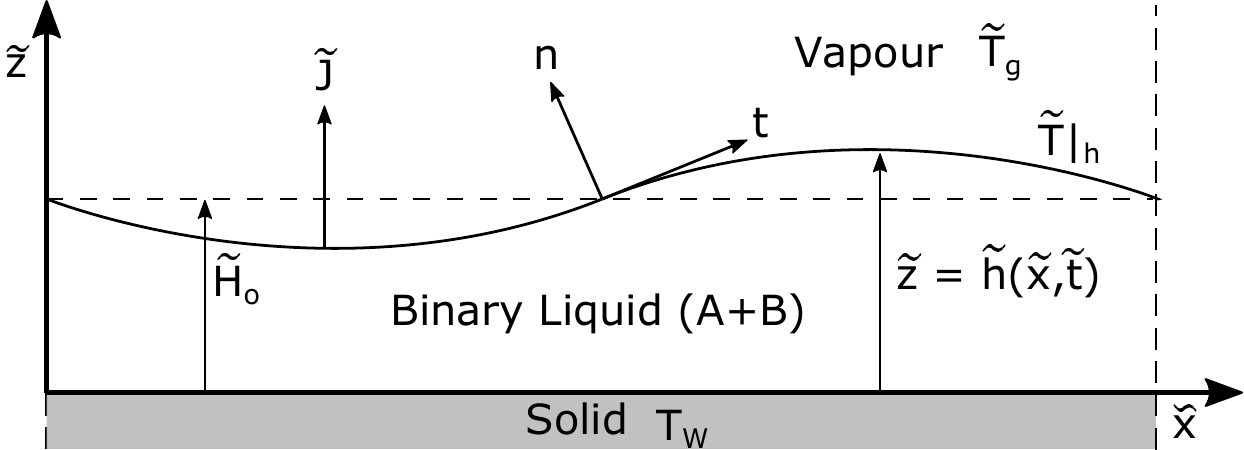}}
	\caption{\small Schematic of the physical system describing an evaporating thin liquid film on top of a horizontal heated solid substrate in a periodic domain, with periodic boundary condition at $\tilde{x} = 0$ and $\tilde{x} = \tilde{L}_o$.}
	\label{fig:section2_schematic}
\end{figure}

The Cartesian coordinate system, $(\tilde{x}, \tilde{z})$, is used to model the dynamics and solve for the velocity field, $\mathbf{\tilde{u}} = (\tilde{u}, \tilde{w})$, where $\tilde{u}$ and $\tilde{w}$ correspond to the horizontal and vertical components of the velocity field, respectively. The temperature field is represented by $\tilde{T}$ and the volume fraction of component A is represented by $c$; since we deal with a binary mixture the volume fraction of component B will be given by $1-c$. The liquid-gas interface is located at $\tilde{z} = \tilde{h}(\tilde{x}, \tilde{t})$ whereas the liquid-solid interface is located at $\tilde{z} = 0$. A sketch of the physical system is presented in Fig. \ref{fig:section2_schematic}.

The flow is incompressible and governed by the mass, momentum, energy and volume fraction conservation equations given by,
\begin{gather}
\tilde{u}_{\tilde{x}} + \tilde{w}_{\tilde{z}} = 0 \\
\tilde{\rho} ( \tilde{u}_{\tilde{t}} + \tilde{u} \tilde{u}_{\tilde{x}} + \tilde{w} \tilde{u}_{\tilde{z}} ) =  - \tilde{p}_{\tilde{x}} 
+ (\tilde{\mu}\tilde{u}_{\tilde{x}})_{\tilde{x}} + (\tilde{\mu}\tilde{u}_{\tilde{z}})_{\tilde{z}} \\
\tilde{\rho} ( \tilde{w}_{\tilde{t}} + \tilde{u} \tilde{w}_{\tilde{x}} + \tilde{w} \tilde{w}_{\tilde{z}} ) =  - \tilde{p}_{\tilde{z}} 
+ (\tilde{\mu}\tilde{w}_{\tilde{x}})_{\tilde{x}} + (\tilde{\mu}\tilde{w}_{\tilde{z}})_{\tilde{z}} \\
\tilde{\rho} ( (\tilde{c}_p \tilde{T})_{\tilde{t}} + \tilde{u} (\tilde{c}_p \tilde{T})_{\tilde{x}} + \tilde{w} (\tilde{c}_p \tilde{T})_{\tilde{z}} ) =  ( \tilde{\lambda} \tilde{T}_{\tilde{x}} )_{\tilde{x}} + ( \tilde{\lambda} \tilde{T}_{\tilde{z}} )_{\tilde{z}} \\
c_{\tilde{t}} + \tilde{u} c_{\tilde{x}} + \tilde{w} c_{\tilde{z}} = \tilde{D}_A (c_{\tilde{x}\tilde{x}} + c_{\tilde{z}\tilde{z}})
\end{gather}

\noindent
where $\tilde{p}$ is the pressure and $D_i$ is the diffusion coefficient of component $i$ $(i = A, B)$. The subscripts $\tilde{x}$, $\tilde{z}$ and $\tilde{t}$ denote spatial and temporal partial differentiation, respectively.

The dependence of the properties of the liquid layer on the local volume fraction of the two components can be evaluated using the following rule of mixtures,
\begin{equation}
\tilde{\phi} = c\tilde{\phi}_A + (1-c)\tilde{\phi}_B 
\end{equation}
where $\tilde{\phi} = \tilde{\mu}, \tilde{\lambda}$ and $\tilde{c}_p$. The density is considered to be constant, $\rho \approx \rho_A \approx \rho_B$, under the assumption that the film is very thin, thus the gravitational effects are negligible.

We assume that the density, viscosity, and thermal conductivity in the liquid phase are all much greater than in the vapour phase. Formally, we take the limits,
\begin{equation}
\frac{\tilde{\rho}_v}{\tilde{\rho}} \rightarrow 0, \qquad \frac{\tilde{\mu}_v}{\tilde{\mu}} \rightarrow 0, \qquad \frac{\tilde{\lambda}_v}{\tilde{\lambda}} \rightarrow 0
\end{equation} 

\noindent
Here, $\tilde{\rho}_v$, $\tilde{\lambda}_v$, and $\tilde{\mu}_v$, denote the density, thermal conductivity and viscosity in the gas phase, respectively. However, we retain the vapour density in Eq. \ref{eq:mass_balance}, where it multiplies the vapour velocity, which may be large.

Along the free surface $(\tilde{z} = \tilde{h}(\tilde{x}, \tilde{t}))$, it is necessary to distinguish between the liquid mixture velocity, $\mathbf{\tilde{u}}$ and the velocity of the interface $\mathbf{\tilde{u}}_s = (\tilde{u}_s, \tilde{w}_s)$. If $\tilde{J}$ denotes the total evaporative flux defined as $\tilde{J} = \tilde{J}_A + \tilde{J}_B$ and $\mathbf{n} = (-\tilde{h}_x, 1)/(1 + \tilde{h}_x^2)^{1/2}$ is the outward-pointing unit normal on the interface then,
\begin{equation}
\mathbf{\tilde{u}} = \mathbf{\tilde{u}}_s + \frac{\tilde{J}}{\tilde{\rho}} \mathbf{n}
\end{equation}

\noindent
whilst the tangential components of both velocities, $\mathbf{\tilde{u}}_\tau = \mathbf{\tilde{u}} - (\mathbf{\tilde{u}} \cdot \mathbf{n})\mathbf{n} = \mathbf{\tilde{u}}_s - (\mathbf{\tilde{u}}_s \cdot \mathbf{n}) \mathbf{n}$, are the same. Moreover, at $\tilde{z} = \tilde{h}(\tilde{x}, \tilde{t})$ the velocity field satisfies the local mass, force and energy balance in the liquid and gas phase written as,
\begin{gather}
\tilde{J} = \tilde{\rho} ( \mathbf{\tilde{u}} - \mathbf{\tilde{u}_s} ) \cdot \mathbf{n} = \tilde{\rho}_v ( \mathbf{\tilde{u}_v} - \mathbf{\tilde{u}_s} ) \cdot \mathbf{n} \label{eq:mass_balance} \\
 - \tilde{p} + \mathbf{n} \cdot \underline{\underline{\tilde{\tau}}} \cdot \mathbf{n} = 2 \tilde{H} \tilde{\sigma} - \tilde{\Pi} - \tilde{p}_v 
\label{eq:normal_stress_balance} \\
 \mathbf{n} \cdot \underline{\underline{\tilde{\tau}}} \cdot \mathbf{t} = \tilde{\nabla}_s \tilde{\sigma} \cdot \mathbf{t} 
\label{eq:tangential_stress_balance} \\
\tilde{J_A} \tilde{L}_{v,A} + \tilde{J_B} \tilde{L}_{v,B} + \tilde{\lambda} \tilde{\nabla} \tilde{T} \cdot \mathbf{n}  \nonumber \\
+ \tilde{J}\big[ \tfrac{1}{2} [(\tilde{u}_v - \tilde{u}_s)\cdot \mathbf{n} ]^2 
- \tfrac{1}{2} [(\tilde{u} - \tilde{u}_s)\cdot \mathbf{n} ]^2 \big]  \nonumber \\
+ (\mathbf{\underline{\underline{\tilde{\tau}}}} \cdot \mathbf{n}) \cdot (\mathbf{\tilde{u}} - \mathbf{\tilde{u}}_s) = 0
\label{energy_balance} 
\end{gather}

\noindent
Here $\mathbf{\tilde{u}}_v$ and $\tilde{T}_v$, denote the velocity field and temperature in the gas phase, respectively; $\underline{\underline{\tilde{\tau}}}$ denotes the stress tensor, $\tilde{J}_i$ and $\tilde{L}_{v,i}$ denote the evaporation flux and specific internal latent heat of vaporization, respectively, of component $i$ $(i = A, B)$. Also, $\mathbf{\tilde{t}} = (1, \tilde{h}_x)/(1 + \tilde{h}^2_x)^{1/2}$ denotes the unit tangential vector on the interface, $2\tilde{H}$ is the mean curvature of the free surface and $\tilde{\nabla}_s$ is the surface gradient operator, respectively defined as,
\begin{equation}
2\tilde{H} = - \tilde{\nabla}_s \cdot \mathbf{n}, \quad \tilde{\nabla}_s = (\mathbf{I} - \mathbf{n}\mathbf{n}) \cdot \tilde{\nabla}
\end{equation}
\noindent
$\tilde{\Pi}$ denotes the disjoining pressure, which accounts for the van der Waals attraction, defined as,
\begin{equation}
\tilde{\Pi} = \frac{A}{6\pi\tilde{h}^3}
\end{equation}

\noindent
where $A$ is the Hamaker constant.

Along the moving interface $(\tilde{z} = \tilde{h}(\tilde{x}, \tilde{t}))$ the following boundary condition for the volume fraction is imposed, 
\begin{equation}
- \tilde{D}_A ( \mathbf{n} \cdot \tilde{\nabla} (\tilde{\rho}_A c) )_{z = h} + \tilde{\rho}_A c (\mathbf{\tilde{u}} - \mathbf{\tilde{u}_s}) \cdot \mathbf{n} = \tilde{J}_A
\end{equation}
\noindent
and the kinematic boundary condition,
\begin{gather}
\tilde{f}(x,z,t) = \tilde{z} - \tilde{h}(x,t), \quad \frac{D\tilde{f}}{D \tilde{t}} = 0 \\
\tilde{h}_t + \tilde{u}_s \tilde{h}_x - \tilde{w}_s = 0
\end{gather}

At the liquid-solid interface ($\tilde{z} = 0$) wall conditions are imposed,
\begin{equation}
\tilde{u} = \tilde{w} = 0, \quad \tilde{T} = \tilde{T}_w
\end{equation}

To complete the description, a constitutive equation that describes the dependence of the interfacial tension on the local volume fraction and interfacial temperature is required. To this end, the following constitutive equation is employed,
\begin{equation}
\tilde{\sigma} = c\tilde{\sigma}_A + (1-c)\tilde{\sigma}_B
\label{eq:surface_tension}
\end{equation}

\noindent
which assumes that the surface tension depends on the local volume fraction of the two components. We also assume the following linear dependence on the temperature,
\begin{equation}
\tilde{\sigma}_i = \tilde{\sigma}_{i,o} - \tilde{\gamma}_i(\tilde{T}|_h - \tilde{T}_o) 
\label{eq:surface_tension_b}
\end{equation}

\noindent
Here, $\tilde{\gamma}_i = -\partial \tilde{\sigma}_i / \partial \tilde{T}$ denotes the temperature coefficient of surface tension for the components $i = A, B$ and $\tilde{\sigma}_{i,o}$ is the surface tension of pure component $i = A, B$ at temperature $\tilde{T}_o$; we may assume $\tilde{T}_o = \tilde{T}_g$, where $\tilde{T}_g$ is the equilibrium vapour temperature.

Finally, we need to employ a constitutive equation for the evaporation fluxes, $\tilde{J}_i$. To this end, we employ the Hertz-Knudsen equation for each species which takes the following form,
\begin{equation} \label{eq:constitutive_J1}
\tilde{J}_i = \tilde{\rho}_{v,i} \tilde{L}_{v,i} \bigg( \dfrac{M_i}{ 2\pi R_g \tilde{T}_g^3 } \bigg)^{1/2} (\tilde{T}|_h - \tilde{T}_g)
\end{equation}

This constitutive equation relates the mass flux $\tilde{J}_i$ of component $i = A, B$ at the interface to the local surface temperature $\tilde{T}|_h$, where $M_i$ is the molecular weight and $R_g$ is the universal gas constant.

Assuming that the gas phase is an ideal gas it is possible to express the vapour density in terms of the partial pressure, $\tilde{\rho}_{v,i} = \tilde{p}_{v,i} M_i / R_g \tilde{T}_g$. Moreover, using Raoult's law the partial pressure can be related to the volume fraction of each component, $\tilde{p}_{v,i} = c_i \tilde{p}_i^o$, where $\tilde{p}_i^o$ is the vapour pressure of component $i = A, B$. Using these relationships the evaporation fluxes can be expressed as,
\begin{equation} \label{eq:constitutive_J2}
\tilde{J}_i = c_i \tilde{p}_i^o \tilde{L}_{v,i} \bigg( \dfrac{M_i^3}{ 2\pi R_g^3 \tilde{T}_g^5 } \bigg)^{1/2} (\tilde{T}|_h - \tilde{T}_g)
\end{equation}

\section{Scaling}

For non-dimensionalising this problem, length is scaled by the initial mean film thickness $\tilde{H}_o$, the viscous scales are used for velocity, time and pressure, the equilibrium vapour temperature $\tilde{T}_g$ is taken as the reference temperature and the properties of component A are taken as reference. The resulting scaling reads,
\begin{gather}
(\tilde{x}, \tilde{z}) =\tilde{H}_o( x, z), 
\quad (\tilde{u}, \tilde{w}) = \dfrac{\tilde{\nu}_A}{\tilde{H}_o} (u, w), \nonumber \\
\quad \tilde{t} = \dfrac{\tilde{H}_o^2}{\tilde{\nu}_A} t, 
\quad \tilde{p} = \dfrac{\tilde{\rho}_A \tilde{\nu}_A^2}{\tilde{H}_o^2} p, 
\quad \tilde{h} = \tilde{H}_o h, \nonumber \\
\quad \tilde{T} = \tilde{T}_g + \Delta \tilde{T} T, 
\quad \tilde{J}_i = \dfrac{\tilde{\lambda}_A \Delta \tilde{T}}{\tilde{H}_o \tilde{L}_{vA}}J_i, \nonumber \\
\quad (\tilde{\tau}_{xz}, \tilde{\tau}_{ii}) = \dfrac{\tilde{\rho}_A \tilde{\nu}_A^2}{\tilde{H}_o^2} ( \tau_{xz}, \tau_{ii}), 
\quad \Delta \tilde{T} = \tilde{T}_w - \tilde{T}_g, \nonumber \\
\quad \tilde{\rho} \approx \tilde{\rho}_A \approx \tilde{\rho}_B, 
\quad \tilde{\mu} = \tilde{\mu}_A \mu, 
\quad \tilde{c}_p = \tilde{c}_{pA} c_p, \nonumber \\
\quad \tilde{\lambda} = \tilde{\lambda}_A \lambda, 
\quad \tilde{\sigma} = \tilde{\sigma}_{A,o} \sigma \nonumber
\end{gather}

\noindent
here $\tilde{\nu}_A$ is the kinematic viscosity of component A.

This scaling renders the following non-dimensional system of governing equations,
\begin{gather}
u_x + w_z = 0 \\
 u_t + uu_x +wu_z = -p_x + (\mu u_x)_x + (\mu u_z)_z \\
 w_t + uw_x + ww_z = -p_z + (\mu w_x)_x + (\mu w_z)_z \\
Pr [ (c_p T)_t + u (c_pT)_x + w(c_pT)_z ] = (\lambda T_x)_x + (\lambda T_z)_z \\
c_t + u c_x + w c_z = Pe^{-1} ( c_{xx} + c_{zz} ) \label{eq:C}
\end{gather}

\noindent
where $Pr = \tilde{\nu}_A / \tilde{\kappa}_A$ is the Prandtl and $Pe = \tilde{\nu}_A / \tilde{D}_A$ the Peclet numbers. Here $\tilde{\kappa}_A$ is the thermal diffusivity of component $A$.

The properties of the liquid are given by,
\begin{equation}
\phi = c + (1-c)\phi_r 
\end{equation}

\noindent
where $\phi = \mu$, $\lambda$ and $c_p$, and $\phi_r = \mu_r$, $\lambda_r$ and $c_{p,r}$ are the viscosity, thermal conductivity and heat capacity ratios, respectively.

At the interface $z = h(x,t)$, the scaled mass, energy, normal stress and tangential stress balance are given by,
\begin{gather}
EJ = ( - h_x ( u - u_s ) + w - w_s ) (h_x^2 + 1)^{-\frac{1}{2}} \label{eq:mass_ibc} \\
J_A + \Lambda J_B + \dfrac{E^2}{2\mathcal{L}D^2}J^3 + \lambda ( -h_x T_x + T_z) (h_x + 1)^{-\frac{1}{2}} = 0 \label{eq:energy_ibc} \\
p - \mathbf{n} \cdot \underline{\underline{\tau}} \cdot \mathbf{n} = p_v + \dfrac{ \mathcal{A} } {h^3 } \nonumber \\ - \bigg(\dfrac{\delta}{Ca} + \dfrac{M_c c}{Pr} - \dfrac{M_T}{Pr}(\gamma_r + (1-\gamma_r)c)T \bigg) \dfrac{h_{xx}}{(h_x^2 + 1)^{\frac{3}{2}}} 
\label{eq:normal_stress_ibc} \\
 \mathbf{n} \cdot \underline{\underline{\tau}} \cdot \mathbf{t} = Ca^{-1} (h_x^2 + 1)^{\frac{1}{2}} (\sigma_x + h_x \sigma_z) 
\label{eq:tangential_stress_ibc} 
\end{gather}

\noindent
here $J = J_A + J_B$ is the total mass flux, $E = \tilde{\lambda}_A \Delta \tilde{T} / \tilde{\rho}_A \tilde{\nu}_A \tilde{L}_{vA}$ is the so-called non-dimensional evaporation number characterizing the evaporation rate, $\Lambda = \tilde{L}_{vB} / \tilde{L}_{vA}$ is the latent heat ratio of the components, $D = \tilde{\rho}_v/\tilde{\rho}$ is the ratio of vapour to liquid densities,  $Ca = \tilde{\rho}_A \tilde{\nu}_A^2 / \tilde{\sigma}_{A,o} H_o$ is the capillary number, $M_c = (\tilde{\sigma}_A - \tilde{\sigma}_B) \tilde{H}_o/\tilde{\rho}_A\tilde{\nu}_A\tilde{\kappa}_A$ is the solutal Marangoni number, $M_T = \tilde{\gamma}_A\Delta\tilde{T}\tilde{H}_o/\tilde{\rho}_A\tilde{\nu}_A\tilde{\kappa}_A$ is the thermal Marangoni number, $\mathcal{L} = \tilde{H}_o^2 \tilde{L}_{vA} / \tilde{\nu}_A^2$ is a measure of the latent heat of component A, and $\mathcal{A} = A / 6\pi \tilde{\rho}_A \tilde{\nu}_A^2 \tilde{H}_o$ is the non-dimensional Hamaker constant $A$.

The scaled boundary condition for the volume fraction reads,
\begin{equation}
\frac{1}{Pe} \bigg[ \frac{- h_x c_x + c_z}{(h_x^2 + 1)^{\frac{1}{2}}} \bigg]_{z = h} = E (cJ - J_A)
\end{equation}

The scaled kinematic boundary condition is given by,
\begin{equation}
h_t + u h_x - w_s = 0
\end{equation}

Using the kinematic boundary condition the mass balance at the interface reads,
\begin{equation}
EJ = (w - h_t - uh_x)(h_x^2 + 1)^{-\frac{1}{2}}
\end{equation}

The scaled constitutive equation for the evaporative flux J is written as,
\begin{align}
KJ_A &= c T \label{eq:ND_JA}\\
KJ_B &= (1-c)\alpha \beta^{\frac{3}{2}} \Lambda T \label{eq:ND_JB}
\end{align}

\noindent
Here, $\alpha = \tilde{p}_B^o/\tilde{p}_A^o$ is the relative volatility, where $\tilde{p}_i^o$ is the vapour pressure of component $i = A, B$ and $\beta = M_B / M_A$ is the molar ratio of the components. The parameter $K$ measures the degree of non-equilibrium at the evaporating interface and is defined by \cite{Burelbach1988},
\begin{equation}
K = \dfrac{\tilde{\lambda}_A}{\tilde{H}_o \tilde{L}_{vA}^2 \tilde{p}_A^o } \bigg( \frac{2\pi R_g^3 \tilde{T}_g^5}{M_A^3} \bigg)^{\frac{1}{2}}
\end{equation}

\noindent
$K = 0$ corresponds to the quasi-equilibrium limit, where the temperature at the interface is constant and equal to the equilibrium vapour temperature, $\tilde{T}_g$. $K \neq 0$ corresponds to the non-equilibrium case and $K^{-1} = 0$ corresponds to the non-volatile case in which the total evaporation flux $J$ is equal to zero.

\begin{table}[!t]
\centering
\caption{\small Properties for water and ethanol at $80\,^\circ C$.}
\begin{tabular}{l*{3}{l}r}
\hline
								& Water 					& Ethanol \\
\hline
$\rho \, [kg\, m^{-3}]$			& $971.82$					& $757$ \\
$\mu \, [N\,s\,m^{-2}]$			& $0.351 \times 10^{-3}$	& $0.432 \times 10^{-3}$ \\
$\lambda\, [W\,m^{-1}\,K^{-1}]$	& $0.669$					& $0.169$ \\
$c_p\, [J\,kg^{-1}\,K^{-1}]$	& $4.197 \times 10^3$		& $3.030 \times 10^3$ \\
$L_v\, [J\,kg^{-1}]$			& $2.309 \times 10^6$		& $0.960 \times 10^6$ \\
$\sigma\, [N\,m^{-1}]$			& $62.69 \times 10^{-3}$	& $17.3 \times 10^{-3}$ \\
$\gamma\, [N\,m^{-1}\,K^{-1}]$	& $0.17 \times 10^{-3} $ 	& $0.09 \times 10^{-3}$ \\
$p^o\, [Pa]$					& $47.37 \times 10^3$		& $108.28 \times 10^3$ \\
$M\, [kg\,mol^{-1}]$			& $18.015 \times 10^{-3}$	& $46.07 \times 10^{-3}$ \\
$D\, [m^2\,s^{-1}]$				& $7.53 \times 10^{-9}$ \\
\hline
\end{tabular}
\label{tab:properties}
\end{table}

The scaled surface tension coefficient is given by,
\begin{equation}
\sigma = c + (1-c)\delta - \Gamma (c + (1 - c)\gamma_r ) T|_h
\label{eq:ND_surface_tension}
\end{equation}

\noindent
where $\delta = \tilde{\sigma}_{B,o} / \tilde{\sigma}_{A,o}$ is the ratio of the reference surface tension, $\gamma_r = \gamma_B / \gamma_A$ is the ratio of the temperature coefficient of surface tension, and $\Gamma = \gamma_A \Delta T / \sigma_A$.

\begin{table}[]
\centering
\caption{\small Dimensionless quantities of a $50\%$ water/ethanol mixture at $80\,^\circ C$ and layer thickness of $1 \, \mu m$.}
\begin{tabular}{l*{4}{l}r}
\hline
$\mathcal{A}$	& $\qquad 1.82\times10^{-5}$		\\
$\alpha$		& $\qquad 2.2858$			\\
$\beta$			& $\qquad 2.5573$			\\
$Ca$			& $\qquad 2.02\times10^{-3}$	\\
$\delta$		& $\qquad 0.276$			\\
$E$ 			& $\qquad 0.0083$			\\
$\Gamma$		& $\qquad 0.0271$			\\
$\gamma_r$		& $\qquad 0.5294$			\\
$K$				& $\qquad 0.1543$				\\
$\mathcal{L}$	& $\qquad 1.77\times10^7$		\\
$\Lambda$		& $\qquad 0.4158$			\\
$\lambda_r$		& $\qquad 0.2526$			\\
$M_C$ 			& $\qquad 7.88\times10^2$		\\
$M_T$			& $\qquad 2.95\times10^{1}$				\\
$\mu_r$			& $\qquad 1.2308$			\\
$Pe$			& $\qquad 47.96$ 			\\
$Pr$			& $\qquad 2.20$ 			\\
\hline
\end{tabular}
\label{tab:parameters1}
\end{table}

Table \ref{tab:properties} shows the properties of water and ethanol at $80 \, ^oC$ and Table \ref{tab:parameters1} shows the corresponding dimensionless quantities for a $50\%$ water/ethanol mixture at $80\,^oC$ and layer thicknesses of $1 \, \mu m$. 

\section{Base state}

In order to perform a linear stability analysis it is necessary to select a base state which will be perturbed with infinitesimal disturbances. As such, we select a film which retains its flat interface as it evaporates slowly, so that it is possible to consider that the base state is quasi-static. Therefore the base state is time-dependent since the heated film is evaporating; the base state quantities will be denoted by an overbar. As a quasi-static base state with a flat evaporating interface is considered, there is no dependence on the lateral coordinate $x$ and the base state velocity field is zero. Since we consider a slowly evaporating film, $E$ is considered to be small and time is rescaled on the evaporative scale. Details of the base state are given in the Appendix A. After expanding the variables in power of E and rescaling time the resulting leading-order base state solution is,
\begin{gather}
\bar{h} = - \dfrac{\bar{\lambda} K}{\bar{\Lambda}_2}+ \dfrac{1}{\bar{\Lambda}_2}\sqrt{ (\bar{\lambda} K + \bar{\Lambda}_2)^2 - 2\bar{\lambda} \bar{\Lambda}_1\bar{\Lambda}_2 Et}
\label{eq:BS_h} \\
\bar{T} = 1 - \dfrac{\bar{\Lambda}_2 z}{\sqrt{ (\bar{\lambda} K + \bar{\Lambda}_2)^2 - 2\bar{\lambda} \bar{\Lambda}_1 \bar{\Lambda}_2 Et}}
\label{eq:BS_T} \\
\bar{J}_A = \dfrac{\bar{\lambda} \bar{c} }{\sqrt{ (\bar{\lambda} K + \bar{\Lambda}_2)^2 - 2\bar{\lambda} \bar{\Lambda}_1 \bar{\Lambda}_2 Et}}
\label{eq:BS_JA} \\
\bar{J}_B = \dfrac{\bar{\lambda} (1 - \bar{c})\alpha\beta^{\frac{3}{2}}\Lambda }{\sqrt{ (\bar{\lambda} K + \bar{\Lambda}_2)^2 - 2\bar{\lambda}  \bar{\Lambda}_1 \bar{\Lambda}_2 Et}}
\label{eq:BS_JB} \\
\bar{p} = A \bigg[\dfrac{\bar{\Lambda}_2}{-\bar{\lambda}K + \sqrt{(\bar{\lambda} K + \bar{\Lambda}_2)^2 - 2\bar{\lambda} \bar{\Lambda}_1 \bar{\Lambda}_2 Et}} \bigg]^3 \label{eq:BS_p} \\
\bar{c}_t = \frac{E\lambda(\Lambda_1-1)\bar{c}}{\bar{h}\sqrt{(\lambda K + \Lambda_2)^2 - 2\lambda\Lambda1\Lambda2} Et}
\label{eq:BS_c_t}
\end{gather}

\noindent
where, $\bar{\Lambda}_1 = \bar{c} + (1- \bar{c})\alpha\beta^{3/2}\Lambda$ and $\bar{\Lambda}_2 = \bar{c} + (1- \bar{c})\alpha\beta^{3/2}\Lambda^2$.

\begin{figure}[!h]
	\centerline{\includegraphics[width=8cm]{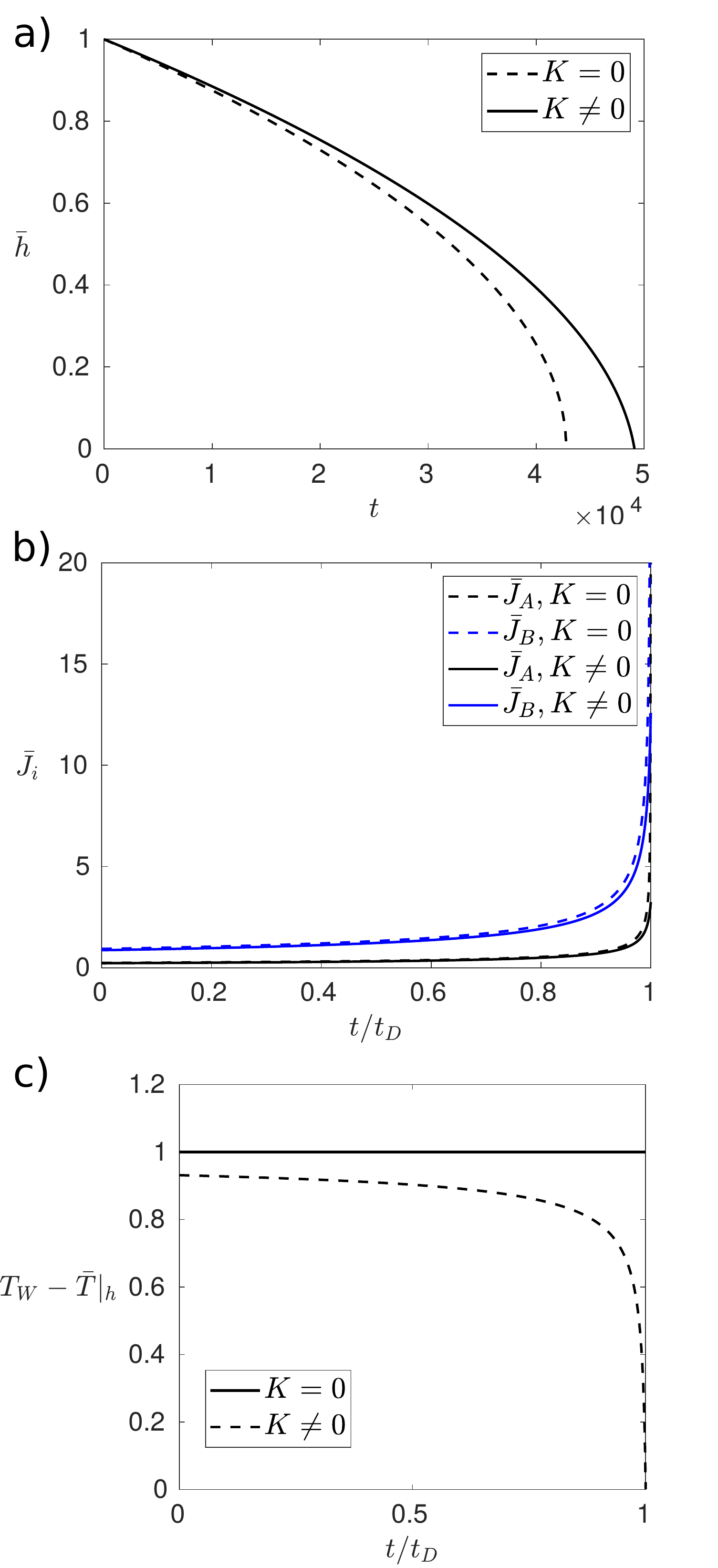}}
	\caption{\small Time evolution of a) the height of the flat interface, b) the evaporation flux of components A and B, and c) the temperature difference between the solid substrate and the interface of the liquid layer, for $K = 0$ and $K \neq 0$. Here $E = 10^{-5}$ and the remaining parameters are shown in Table \ref{tab:parameters1}.}
	\label{fig:section4_base_state}
\end{figure}

Fig. \ref{fig:section4_base_state} shows the time evolution of the basic state quantities. Fig. \ref{fig:section4_base_state}a shows that for $K = 0$ the film thickness goes to zero at $t_D = \frac{\bar{\Lambda}_2}{2\bar{\lambda} \bar{\Lambda}_1 E}$, while for $K \neq 0$ the film disappearance time is $t_D = \frac{2\bar{\lambda} K + \bar{\Lambda}_2}{2 \bar{\lambda} \bar{\Lambda}_1 E}$ that is higher than the quasi-equilibrium case. The evaporative flux of component A and B are shown in Fig. \ref{fig:section4_base_state}b. For $K = 0$ they are initially $\bar{J}_A = \frac{\bar{\lambda} \bar{c} }{\bar{\Lambda}_2}$ and $\bar{J}_B = \frac{\bar{\lambda} (1 - \bar{c})\alpha\beta^{\frac{3}{2}}\Lambda }{\bar{\Lambda}_2}$, respectively, and the most volatile component (component B) has a stronger evaporation flux during the evaporation process. Both go to infinity at the film disappearance time, $t_D$, since for $K=0$ the temperature difference between the interface and the heated substrate is constant during the evaporation process, see Fig. \ref{fig:section4_base_state}c. For $K \neq 0$ the evaporative flux is initially $\bar{J}_A = \frac{\bar{\lambda} \bar{c} }{\bar{\lambda} K + \bar{\Lambda}_2}$ and $\bar{J}_B = \frac{\bar{\lambda} (1 - \bar{c})\alpha\beta^{\frac{3}{2}}\Lambda }{\bar{\lambda} K + \bar{\Lambda}_2}$ and increase to $\bar{J}_A = \frac{\bar{c} }{K}$ and $\bar{J}_B = \frac{(1 - \bar{c})\alpha\beta^{\frac{3}{2}}\Lambda }{K}$ at the disappearance time. Fig. \ref{fig:section4_base_state}c shows that the temperature difference between the interface and the solid substrate is constant for $K = 0$ and for $K \neq 0$ it is initially $T_W - \bar{T}|_h = \frac{\bar{\Lambda}_2}{\bar{\lambda} K + \bar{\Lambda}_2}$ and decreases to zero as the height of the film becomes smaller and the temperature at the interface approaches the substrate temperature.

\section{Long-wave approach}

Assuming that the evaporation is a slow process and that the horizontal extent of the liquid layer is much larger than the vertical extent, the long-wave approximation can be applied. To that end the governing system of equations cen be rescaled using the small parameter $\varepsilon = \tilde{H}_o/\tilde{L}_o$ by writing,
\begin{equation}
X = \varepsilon x, \quad Z = z, \quad \tau = \varepsilon t
\end{equation}

We assume that $u, \, J_A,\, J_B,\, T$ are O(1) while $w$ is O($\varepsilon$) in order to preserve continuity, and $p$ is of O($\varepsilon^{-1}$). These dependent variables are expanded in powers of $\varepsilon$,
\begin{align}
u &= U_o + \varepsilon U_1 + ... \\
w &= \varepsilon (W_o + \varepsilon W_1 + ...) \\
J_A &= J_{Ao} + \varepsilon J_{A1} + ... \\
J_B &= J_{Bo} + \varepsilon J_{B1} + ... \\
T &= T_o + \varepsilon T_1 + ... \\
p &= \varepsilon^{-1} ( P_o + \varepsilon P_1 + ...)
\end{align}

For the volume fraction profile in the vertical direction we consider the rapid diffusion approximation \cite{Jensen1993,Warner2003,Craster2009b}, in which the volume fraction, $c$, is decomposed into a $z$-averaged component and a small perturbation incorporating the $z$ dependence,
\begin{equation}
c(X,Z,\tau) = C_{o}(X,\tau) + \varepsilon^2 Pe^* C_{1} (X,Z,\tau)
\end{equation}

\noindent
where $Pe^* = \varepsilon^{-1}Pe$. Since, $\varepsilon^2Pe^* \ll 1$, the decomposition of $c$ allows us to consider a limit in which the vertical volume fraction gradients are negligible. In the small $\varepsilon$ limit, we have the following leading order system of governing equations,
\begin{gather}
U_{o,X} + W_{o,Z} = 0 \label{eq:LW_mass} \\
P_{o,X} = (\mu U_{o,Z})_Z \label{eq:LW_mom_x} \\
P_{o,Z} = 0 \label{eq:LW_mom_z} \\
(\lambda T_{o,Z})_Z = 0 \label{eq:LW_energy} \\
C_{o,\tau} + U_o C_{o,X} = \frac{1}{Pe^*} C_{o,XX} + C_{1,ZZ} \label{eq:LW_concentration}
\end{gather}

At the wall ($Z = 0$), we have,
\begin{equation}
U_o = W_o = 0, \quad T_o = 1 
\label{eq:LW_z0}
\end{equation}

Along the interface $(Z = h(X,\tau))$, the boundary conditions become,
\begin{gather}
E^* J_o = W_o - h_{\tau} - U_o h_X 
\label{eq:LW_kinematic_BC} \\
J_{o,A} + \Lambda J_{o,B} = - \lambda T_{o,Z} \label{eq:LW_energy_balance} \\
P_o = p_v - \frac{\delta}{Ca^*} h_{XX} + \frac{\mathcal{A}^*}{h^3} 
\label{eq:LW_normal_stress} \\
\mu U_{o,Z} = \frac{M_c^* C_{o,X}}{Pr} \nonumber \\
- \frac{M_T}{Pr}^* [ (1-\gamma_r)C_{o,X} T|_h + (C_o + (1-C_o)\gamma_r) T_X|_h]] 
\label{eq:LW_tangential_stress}  \\
C_{1,Z} |_{Z = h} = E^* (C_o J_o - J_{o,A}) + \frac{h_X C_{o,X}}{Pe^*} 
\label{eq:LW_BC_concentration}
\end{gather}

\noindent
Since $E$ is considered to be small we assume $E^* = \varepsilon^{-1}E$ to include mass loss in the kinematic boundary condition. The kinetic energy in the energy balance is neglected by assuming $\mathcal{L} = O(\varepsilon^5)$. We assume $M_T^* = \varepsilon^{-1} M_T$ and $M_c^* = \varepsilon^{-1} M_c$ to retain the thermocapillary and solutal effect in the tangential stress balance. We also assume $Ca^* = \varepsilon^{-3} Ca$ and $\mathcal{A}^* = \varepsilon \mathcal{A}$ to retain the effect of surface tension and disjoining pressure in the normal stress balance, respectively.

The constitutive equation for evaporation flux at the leading order reads,
\begin{align}
K J_{o,A} &= C_o T_o|_h \label{eq:LW_JA} \\
K J_{o,B} &= (1 - C_o) \alpha\beta^{\frac{3}{2}}\Lambda T_o|_h \label{eq:LW_JB}
\end{align}

The leading order surface tension coefficient is given by,
\begin{equation}
\sigma = C_o + (1 - C_o) \delta -  \Gamma_A ( C_o +  (1 - C_o) \gamma_r ) T_o|_h
\label{eq:LW_surf_tens}
\end{equation}

First, we solve the energy conservation Eq. (\ref{eq:LW_energy}), subject to the energy balance Eq. (\ref{eq:LW_energy_balance}) and the wall boundary conditions Eq. (\ref{eq:LW_z0}) to find the liquid temperature field,
\begin{equation}
T_o = 1 - (J_{o,A} + \Lambda J_{o,B}) \frac{Z}{\lambda}
\label{eq:LW_To}
\end{equation}

To find the velocity, we solve $x$-component of the conservation of momentum Eq. (\ref{eq:LW_mom_x}), subject to the tangential stress balance Eq. (\ref{eq:LW_tangential_stress}), and the wall boundary condition Eq. (\ref{eq:LW_z0}),
\begin{gather}
U_o = \frac{P_{o,X}}{\mu} \Big( \frac{Z^2}{2} - hZ \Big) 
+ \frac{\sigma_x Z}{\mu Ca}
\label{eq:LW_Uo}
\end{gather}

From the conservation of mass Eq. (\ref{eq:LW_mass}), subject to the wall boundary conditions Eq. (\ref{eq:LW_z0}), we have,
\begin{gather}
W_o = - \frac{P_{o,XX}}{\mu} \Big( \frac{Z^3}{6} - \frac{hZ^2}{2} \Big) + \frac{P_{o,X} h_X Z^2}{2\mu} - \frac{\sigma_{XX} Z^2}{2\mu Ca}
\label{eq:LW_Wo}
\end{gather}

From the kinematic boundary condition Eq. (\ref{eq:LW_kinematic_BC}), and integrating over $z$ the conservation of species Eq. (\ref{eq:LW_concentration}), together with the boundary condition for the volume fraction Eq. (\ref{eq:LW_BC_concentration}), we have the following evolution equations,
\begin{gather}
h_{\tau} = - E^* J_o + \Big( \frac{P_{o,X} h^3}{3\mu} 
- \frac{M_c^* C_{o,X} h^2}{2\mu Pr} \nonumber \\
+ \frac{M_T^* [ (1-\gamma_r)C_{o,X} T|_h - \gamma_o T_X|_h] h^2 }{2\mu Pr} \Big)_X 
\label{eq:LW_h_t}
\end{gather}
\begin{gather}
C_{o,\tau} = \frac{E^* (C_o J_o - J_{o,A})}{h} + \frac{(h C_{o,X})_X}{h Pe^*} \nonumber \\
+ \Big( \frac{P_{o,X} h^2}{3\mu} 
- \frac{ M_c^* C_{o,X} h}{2\mu Pr} \nonumber \\
+ \frac{M_T^* [ (1-\gamma_r)C_{o,X} T|_h - \gamma_o T_X|_h] h}{2\mu Pr} \Big) C_{o,X}
\label{eq:LW_C_t}
\end{gather}

\noindent
where $\gamma_o = C_o + (1-C_o)\gamma_r$.

Returning to the original scaling the evolution equations take the form,
\begin{gather}
h_t = - E J + \Big( \frac{p_x h^3}{3\mu} 
- \frac{M_c C_x h^2}{2\mu Pr} \nonumber \\
+ \frac{M_T [ (1-\gamma_r)C_x T|_h - \gamma_o T_x|_h] h^2 }{2\mu Pr} \Big)_x 
\label{eq:LW_h_t}
\end{gather}
\begin{gather}
C_t = \frac{E (C J - J_A)}{h} + \frac{(h C_x)_x}{h Pe} + \Big( \frac{p_x h^2}{3\mu} - \frac{M_c C_x h}{2\mu Pr} \nonumber \\
+ \frac{M_T [ (1-\gamma_r)C_x T|_h - \gamma_o T_x|_h] h}{2\mu Pr} \Big) C_x
\label{eq:LW_C_t}
\end{gather}

Writing $J$, $p$, and $T$ in terms of $C$ and $h$ and substituting in Eqs. (\ref{eq:LW_h_t}) and (\ref{eq:LW_C_t}) renders the following set of evolution equations written in terms of $C$ and $h$,
\begin{multline}
h_t = - \frac{E \lambda \Lambda_1}{\lambda K + \Lambda_2 h} + \bigg[ - \frac{\delta h^3 h_{xxx}}{3\mu Ca} 
- \frac{\mathcal{A} h_x}{\mu h} \\
- \frac{M_c C_xh^2}{2\mu Pr}  
- \frac{M_T h^2}{2\mu Pr} \bigg( (1 - \gamma_r) C_x \bigg( \frac{\lambda K}{\lambda K + \Lambda_2 h} \bigg) \\
- \gamma_o \bigg( \frac{\lambda K (\Lambda_{2,x} h + \Lambda_2 h_x)}{(\lambda K + \Lambda_2 h)^2} \bigg) \bigg) \bigg]_x
\label{eq:LW_h_t2}
\end{multline}
\begin{multline}
C_t = \frac{E\lambda(\Lambda_1 - 1) C}{h(\lambda K + \Lambda_2 h)} 
+ \bigg[ - \frac{\delta h^2 h_{xxx}}{3\mu Ca} 
- \frac{\mathcal{A} h_x}{\mu h^2} \\
- \frac{M_c C_x h}{2\mu Pr} 
+ \frac{M_T h}{2\mu Pr} \bigg( (1 - \gamma_r) C_x \bigg( \frac{\lambda K}{\lambda K + \Lambda_2 h} \bigg) \\
- \gamma_o \bigg( \frac{\lambda K (\Lambda_{2,x} h + \Lambda_2 h_x)}{(\lambda K + \Lambda_2 h)^2} \bigg) \bigg) \bigg] C_x 
+ \frac{(h C_x)_x}{h Pe}
\label{eq:LW_C_t2}
\end{multline}

\noindent
where $\Lambda_1 = C + (1-C)\alpha\beta^{3/2}\Lambda$ and $\Lambda_2 = C + (1-C)\alpha\beta^{3/2}\Lambda^2$.

\section{Linear stability analysis}

\begin{table*}
\caption{\small The expressions and orders of magnitude of the terms in Eqs. \ref{eq:LW_h_new} and \ref{eq:LW_C_new}, where $\mu_b = C_b + (1 - C_b)\mu_r$, and $\gamma_b = C_b + (1-C_b)\gamma_r$.}
\begin{ruledtabular}
\begin{tabular}{l*{3}{l}r}
Physics						& Jacobian 										& Order \\
\hline
Solutal Marangoni				& $M_c^{HC} = \frac{M_c h_b^2}{2 \mu_b Pr} $					& $\mathcal{O}(10^{2})$ \\

Surface tension				& $S^{HH} = \frac{\delta h_b^3 }{3\mu_b Ca}$					& $\mathcal{O}(10^{1})$ \\

Thermal Marangoni				& $M_T^{HH} = \frac{M_T h_b^2 \lambda_b K \gamma_b \Lambda_{2b}}{2\mu_b Pr(\lambda_b K + \Lambda_{2b}h_b)^2}$					& $\mathcal{O}(10^{-1})$ \\

Thermal Marangoni				& $M_T^{HC} = \frac{M_T h_b^2\lambda_b K}{2 \mu_b Pr(\lambda_b K + \Lambda_{2b} h_b)} \bigg( (1 - \gamma_r) - \frac{\gamma_b h_b (1 - \alpha\beta^{3/2}\Lambda^2)}{(\lambda_b K + \Lambda_{2b} h_b)} \bigg)$					& $\mathcal{O}(10^{-1})$ \\

Diffusion volume fraction			& $D^{CC} =  \frac{1}{Pe}$							& $\mathcal{O}(10^{-2})$ \\

Disjoining pressure				& $\mathcal{A}^{HH} = \frac{\mathcal{A} }{\mu_b h_b}$				& $\mathcal{O}(10^{-5})$ \\

Evaporation					& $E^{HH} = \frac{E \lambda_b \Lambda_{1b} \Lambda_{2b}}{(\lambda_b K + \Lambda_{2b} h_b)^2}$																				& $\mathcal{O}(10^{-5})$ \\

Evaporation					& $E^{HC} = \frac{E \lambda_b ((1 - \alpha\beta^{3/2}\Lambda)}{(\lambda_b K + \Lambda_{2b} h_b)} 
+ \frac{E\Lambda_{1b}h_b ((1-\lambda_r)\Lambda_{2b} - \lambda_b(1-\alpha\beta^{3/2}\Lambda^2))}{(\lambda_b K + \Lambda_{2b} h_b)^2} $								& $\mathcal{O}(10^{-6})$ \\

Evaporation					& $E^{CH} =  \frac{E \lambda_b (\Lambda_{1b} - 1)C_b}{h_b(\lambda_b K + \Lambda_{2b} h_b)} \bigg( \frac{1}{h_b} + \frac{\Lambda_{2b}}{\lambda_b K + \Lambda_{2b} h_b} \bigg)$
																	& $\mathcal{O}(10^{-6})$ \\

Evaporation					& $E^{CC} =  \frac{E((1-\lambda_r)(\Lambda_{1b}-1)C_b + \lambda_b(1-\alpha\beta^{3/2}\Lambda)C_b + \lambda_b(\Lambda_{1b}-1))}{h_b(\lambda_b K + 
\Lambda_{2b} h_b)} - \frac{E\lambda_b(\Lambda_{1b}-1)C_b ((1-\lambda_r)K + (1 - \alpha\beta^{3/2}\Lambda^2)h_b)}
{h_b (\lambda_b K + \Lambda_{2b} h_b)^2}$				& $\mathcal{O}(10^{-6})$ \\

\end{tabular}
\end{ruledtabular}
\label{tab:terms_new}
\end{table*}

Considering the linear stability of this state, we perturb the base state in the following form,
\begin{align}
h(x,\tau) &= h_b(\tau) + H(\tau) e^{ikx} \label{eq:LS_h} \\
C(x,\tau) &= C_b + C_1(\tau) e^{ikx} \label{eq:LS_C}
\end{align}

\noindent
where $k$ is the wavenumber. The base solution for $h$ and $C$ are given by Eqs. (\ref{eq:BS_h}) and (\ref{eq:BS_c_t}), and its time derivatives at $t = 0$ can be written as,
\begin{equation}
\dot{h_b} = -\frac{E\lambda_b\Lambda_{1b}}{\lambda_b K + \Lambda_{2b}h_b}, \qquad \dot{C}_b = \frac{E\lambda_b(\Lambda_{1b}-1) C_b}{h_b(\lambda_b K + \Lambda_{2b})}
\end{equation}

\noindent
where $\lambda_b = C_b + (1 - C_b) \lambda_r$, $\Lambda_{1b} = C_b + (1-C_b)\alpha\beta^{3/2}\Lambda$ and $\Lambda_{2b} = C_b + (1-C_b)\alpha\beta^{3/2}\Lambda^2$. 

Substituting the perturbed solution, Eqs. (\ref{eq:LS_h}) and (\ref{eq:LS_C}) into the system of equations, Eqs. (\ref{eq:LW_h_t2}) and (\ref{eq:LW_C_t2}) and linearising these with respect to $H$ and $C_1$, we obtain the following linear system,
\begin{multline}
\dot{H} + \bigg[ - E^{HH} - ( \mathcal{A}^{HH} + M_T^{HH}) k^2 + S^{HH}k^4  \bigg] H \\
+ \bigg[ - E^{HC} + ( M_T^{HC} - M_c^{HC} ) k^2 \bigg] C_1 = 0
\label{eq:LW_h_new}
\end{multline}
\begin{equation}
\dot{C_1} + \bigg[ E^{CH} \bigg] H + \bigg[ - E^{CC} + \mathcal{D}^{CC} k^2 \bigg] C_1 = 0
\label{eq:LW_C_new}  
\end{equation}
\noindent
where the Jacobian terms, $\Phi^{ij}$, and their orders of magnitude are given in Table \ref{tab:terms_new}. In the Jacobian terms, the superscript $i = H, C$ refers to the terms in the interfacial and volume fraction equations, respectively, and the superscript $j = H, C$ indicates if the Jacobian term is multiplying $H$ or $C$, respectively.

Below we use the `frozen' interface approximation \cite{Burelbach1988, MikishevNepomnyashchy2013}, which assumes that the characteristic time of the change of the layer thickness is large compared to the development of the disturbances. This allows us to disregard the dependence of $h_b$ on $\tau$ considering it as a constant parameter. In that case, we consider the following disturbances,
\begin{equation}
H(\tau) = H(0) e^{r\tau}, \quad C_1(\tau) = C_1(0) e^{r\tau}
\end{equation}

\noindent
where $r$ denotes the growth rate of the disturbances and $H(0)$, $C_1(0)$ the imposed disturbance. With that we obtain the following set of equations,
\begin{multline}
\bigg[r - E^{HH} - ( \mathcal{A}^{HH} + M_T^{HH}) k^2 + S^{HH}k^4  \bigg] H \\
+ \bigg[ - E^{HC} + ( M_T^{HC} - M_c^{HC}) k^2 \bigg] C_1 = 0
\label{eq:LSA_H}
\end{multline}
\begin{equation}
\bigg[ E^{CH} \bigg] H + \bigg[ r - E^{CC} + \mathcal{D}^{CC} k^2 \bigg] C_1 = 0
\label{eq:LSA_C}  
\end{equation}

Next, we will solve the Eqs. \ref{eq:LSA_H} and \ref{eq:LSA_C} to get an expression for the growth rate $r_\pm$, as a function of the wavenumber $k$ for different cases. The solution of $r_\pm$ has two possibilities: 
\begin{enumerate}[label=(\alph*)]
\item Two real roots that correspond to the monotonic damping or growth (depending on the sign of the root) of the disturbances. The two roots correspond to the growth rate of the two different modes, i.e. the interfacial and the volume fraction mode. 
\item Two complex roots that correspond to an oscillatory mode of instability. The real part of $r$ gives the growth rate while the imaginary part the frequency of the instability.
\end{enumerate}

\subsection{Quasi-equilibrium evaporation ($K = 0$)}

We consider first the quasi-equilibrium case, where the interfacial temperature is constant and equal to the equilibrium temperature. Under this condition, the thermocapillary effect is absent. For the case without evaporation the growth rate is given by the interfacial and volume fraction modes respectively,
\begin{align}
r_+ &= \mathcal{A}^{HH}k^2 - S^{HH}k^4 \\
r_- &= - D^{CC} k^2
\end{align}

\begin{figure}[!t]
	\centerline{\includegraphics[width=7cm]{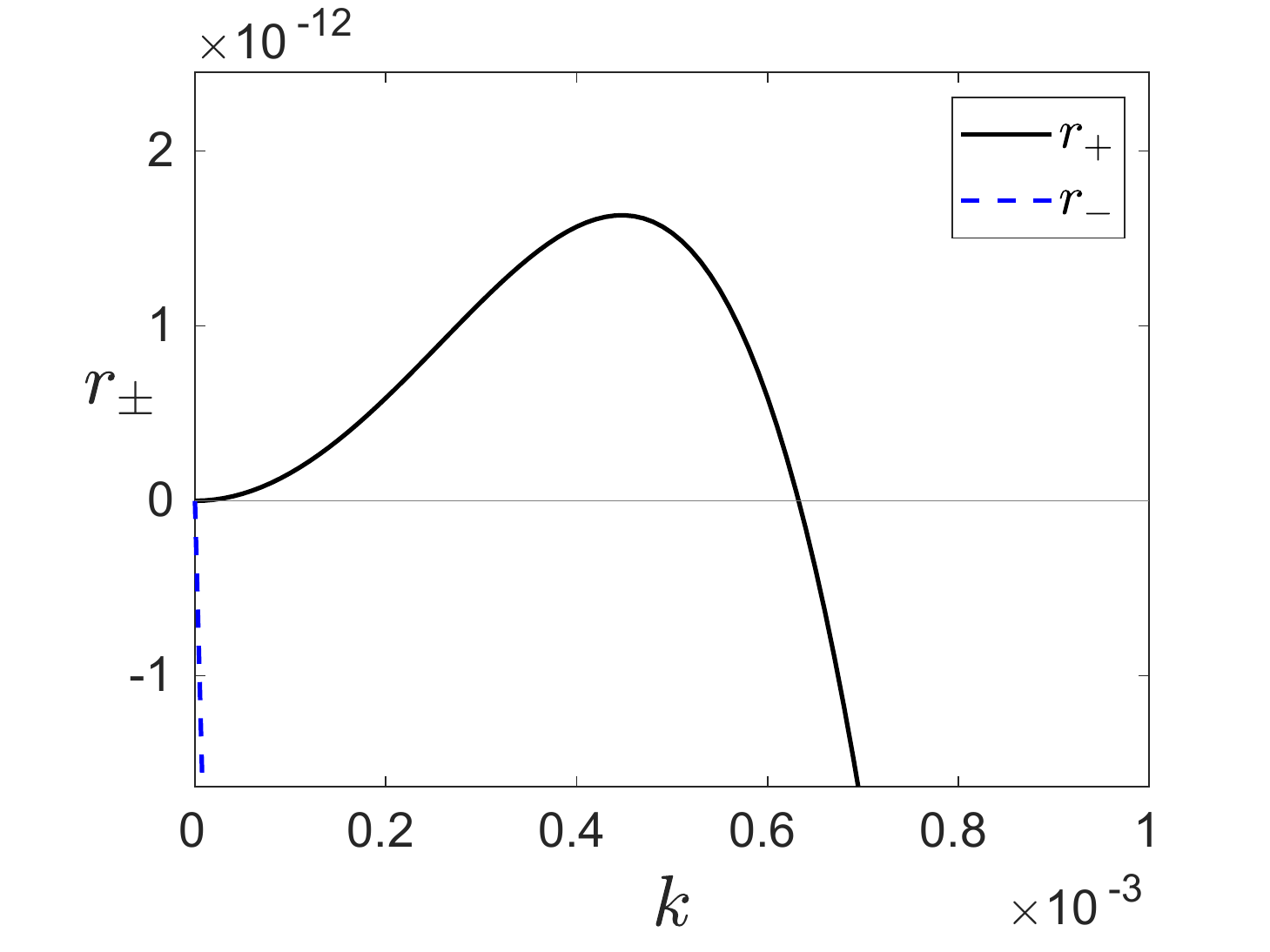}}
	\caption{\small The growth rates $r_\pm$ versus wavenumber $k$, for $K = 0$ and $E = 0$. The remaining parameters are shown in Table \ref{tab:parameters1}.}
	\label{fig:section6A_growth_rate_K0_E0}
\end{figure}

Fig. \ref{fig:section6A_growth_rate_K0_E0} shows that without evaporation the interfacial mode is unstable while the volume fraction mode is stable. In this case the dominant effects in the interfacial mode are the van der Waals attractions that destabilise the layer at very small wavenumbers and the surface tension stabilise the layer at large wavenumbers, while for the volume fraction mode the diffusion of components of the mixture has an stabilising effect.

Next we consider the case with evaporation. For this case the expression for the growth rate, $r\pm$, as a function of the wavenumber, $k$, is given by,
\begin{multline}
r_\pm = \frac{1}{2} \bigg[ E^{HH} + E^{CC} 
+ ( \mathcal{A}^{HH} - \mathcal{D}^{CC}) k^2 - S^{HH}k^4 \bigg] \\
\pm \frac{1}{2} \sqrt{d_1}
\label{eq:growth_rate_K0}
\end{multline}

\noindent
where,
\begin{multline}
d_1 = \bigg(  - E^{HH} +  E^{CC} 
- ( \mathcal{A}^{HH} + \mathcal{D}^{CC}) k^2 + S^{HH}k^4 \bigg)^2 \\
+4 \bigg( E^{CH} \bigg) \bigg(  E^{HC} - M_c^{HC} k^2 \bigg)
\label{eq:discriminant_K0}
\end{multline}

\begin{figure}[!t]
	\centerline{\includegraphics[width=7cm]{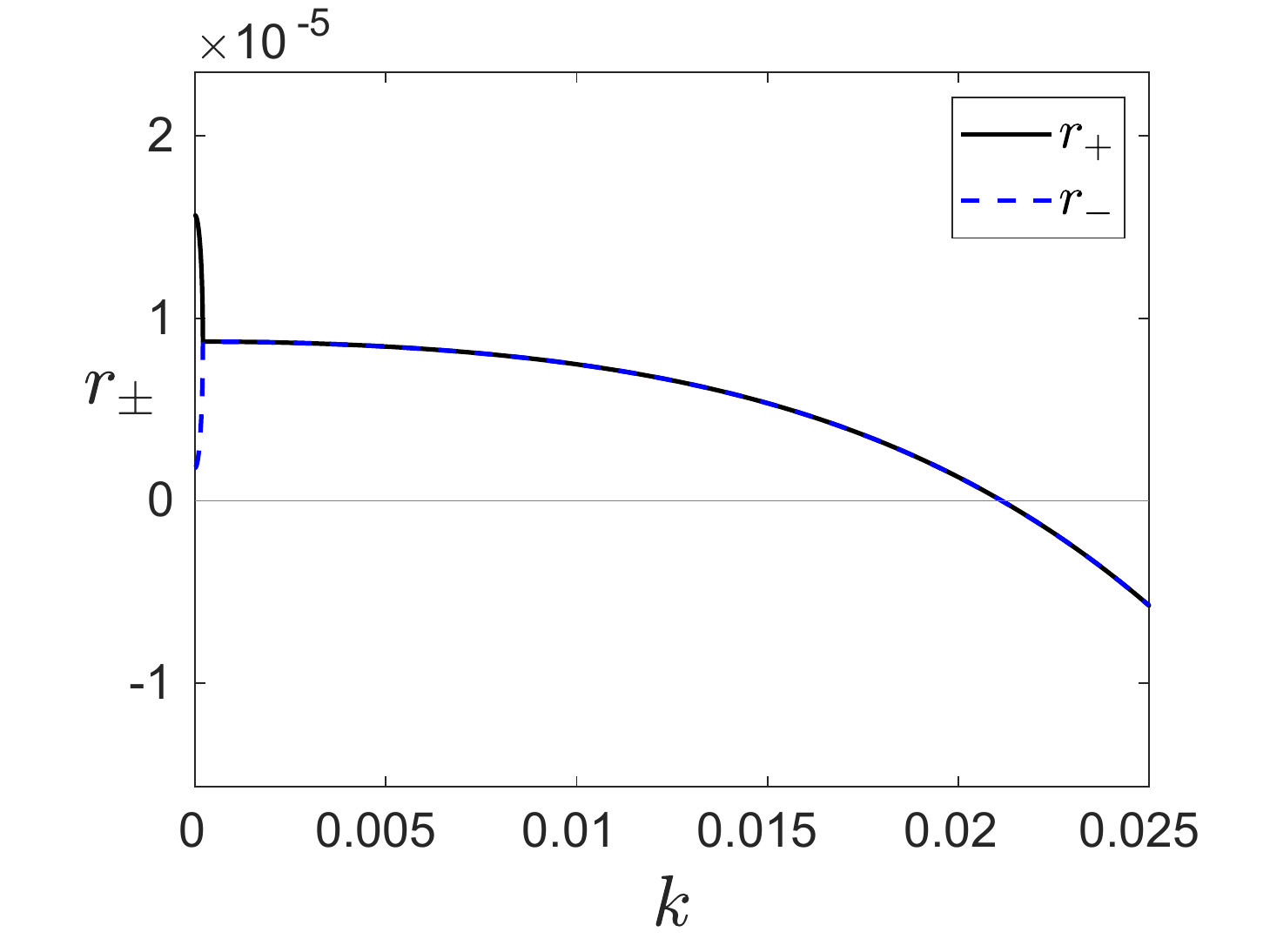}}
	\caption{\small The growth rates $r_\pm$ versus the wavenumber $k$, for $K = 0$ and $E = 10^{-5}$. The remaining parameters are shown in Table \ref{tab:parameters1}.}
	\label{fig:section6A_growth_rate_K0}
\end{figure}

With evaporation, as seen in Fig. \ref{fig:section6A_growth_rate_K0}, the system has real eigenvalues for very small wavenumbers indicating a monotonic instability and complex eigenvalues for moderate wavenumbers indicating an oscillatory instability. As shown in Fig. \ref{fig:section6A_growth_rate_K0} the real part of the complex eigenvalues is of magnitude comparable and lower than the evaporation parameter $E$, thereby suggesting that this oscillatory mode will eventually be overhauled by the long-wave monotonic mode due the evaporation of the film. A decay of the amplitude of the oscillations was also seen in the experiments performed by Overdiep \cite{Overdiep1986}. For this case, evaporation destabilises the liquid layer at small wavenumbers while the diffusion of the mixture components and the surface tension stabilise the liquid layer at high wavenumbers. We can see in the discriminant $d_1$ that the evaporation will make the system monotonic ($d_1 > 0$) at $k = 0$ while the solutal Marangoni effect will make the system oscillatory ($d_1 < 0$) at small wave numbers. The presence of the solutal Marangoni effect due to evaporation reverses the initial perturbation leading the system to an oscillatory instability mode. This case is similar to described by Overdiep \cite{Overdiep1986}, Hovison et al. \cite{Howison1997} and Eres et al. \cite{Eres1999} analysing drying of painting layers, where the increase of resin in the troughs due to evaporation of the solvent increase the surface tension at the troughs and reverses an initial perturbation. In the case of $50\%$ water-ethanol studied in our work, the difference in volatility is the mechanism that will increase the surface tension at the troughs due to the faster evaporation of the most volatile component that has lower surface tension.

\subsection{Non-equilibrium evaporation ($K \neq 0$)}

We now consider the non-equilibrium evaporation, where the interfacial temperature is not constant and depends on the evaporation fluxes. This means that the thermal Marangoni effect is present. For the case without evaporation the interfacial and volume fraction modes are given by respectively,
\begin{align}
r_+ &= (\mathcal{A}^{HH} + M_T^{HH})k^2 - S^{HH}k^4 \\
r_- &= - D^{CC} k^2
\end{align}

\begin{figure}[]
	\centerline{\includegraphics[width=7cm]{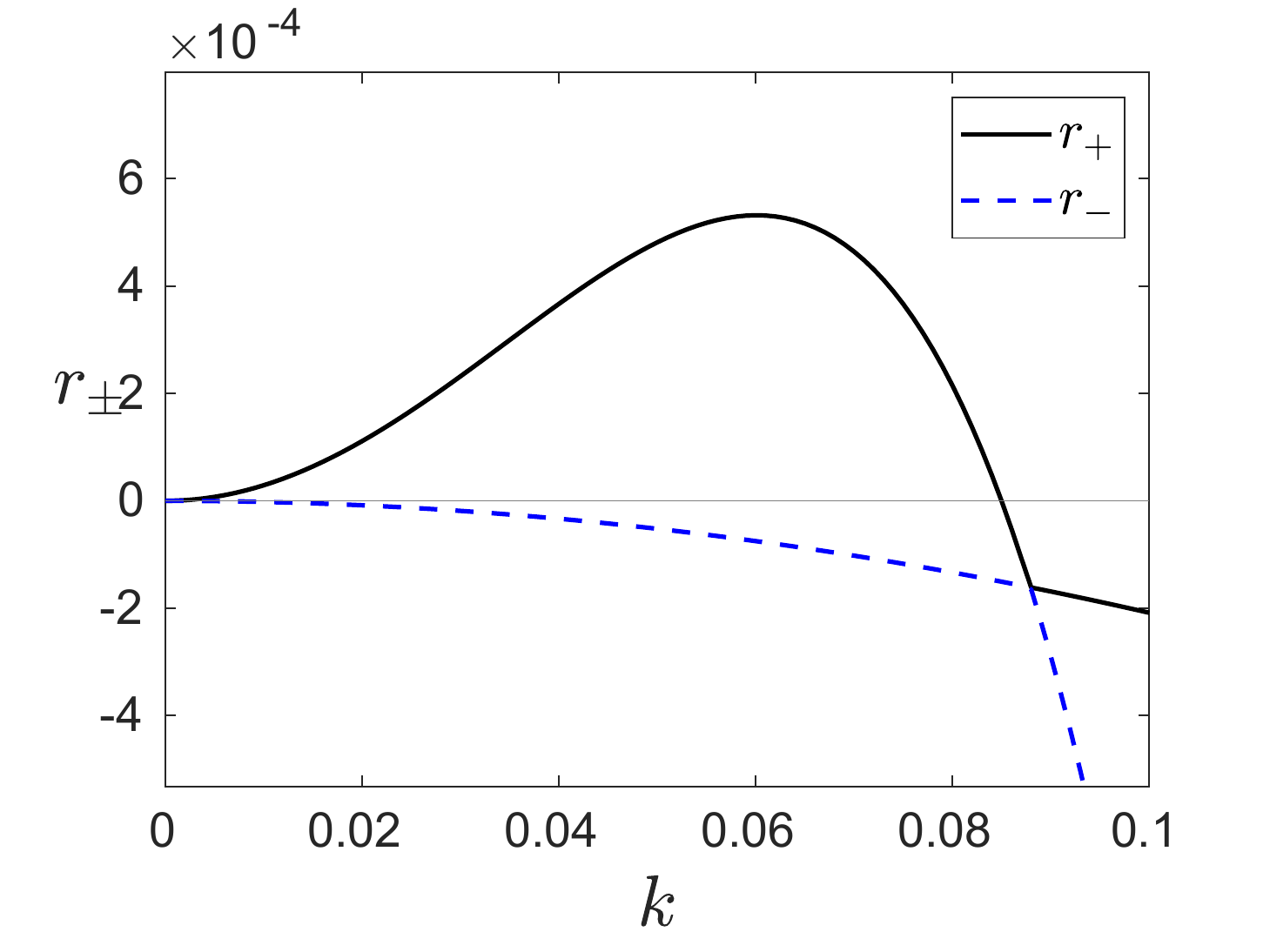}}
	\caption{\small The growth rates $r_+$ and $r_-$ versus the wavenumber k, for $E = 0$. The remaining parameters are shown in Table \ref{tab:parameters1}.}
	\label{fig:section6A_growth_rate_E0}
\end{figure}

Without evaporation, Fig. \ref{fig:section6A_growth_rate_E0} shows that the growth rate, has real eigenvalues and the system is unstable for small wavenumbers. In this case the increase in the temperature at the trough lowers the local surface tension and the thermal Marangoni effect drives the liquid to the crest promoting the initial perturbation leading to a monotonic instability mode. For small wavenumbers the thermal Marangoni effect dominates the instability, while for large wavenumbers the surface tension dominates and stabilises the liquid layer.

Next we consider the case with evaporation where the growth rate $r$ as a function of the wavenumber $k$ is given by,
\begin{multline}
r_\pm = \frac{1}{2} \bigg[ E^{HH} + E^{CC} + ( \mathcal{A}^{HH} + M_T^{HH} - \mathcal{D}^{CC}) k^2 \\
- S^{HH}k^4  \bigg] 
\pm \frac{1}{2} \sqrt{d_1}
\label{eq:growth_rate}
\end{multline}

\noindent
where,
\begin{gather}
d_1 = \bigg(  - E^{HH} +  E^{CC} - ( \mathcal{A}^{HH} \nonumber \\
+ M_T^{HH} + \mathcal{D}^{CC}) k^2 + S^{HH}k^4 \bigg)^2 \nonumber \\
+4 \bigg( E^{CH} \bigg) \bigg(  E^{HC} + ( M_T^{HC} - M_c^{HC} ) k^2 \bigg)
\label{eq:discriminant}
\end{gather}

\begin{figure}[]
	\centerline{\includegraphics[width=7cm]{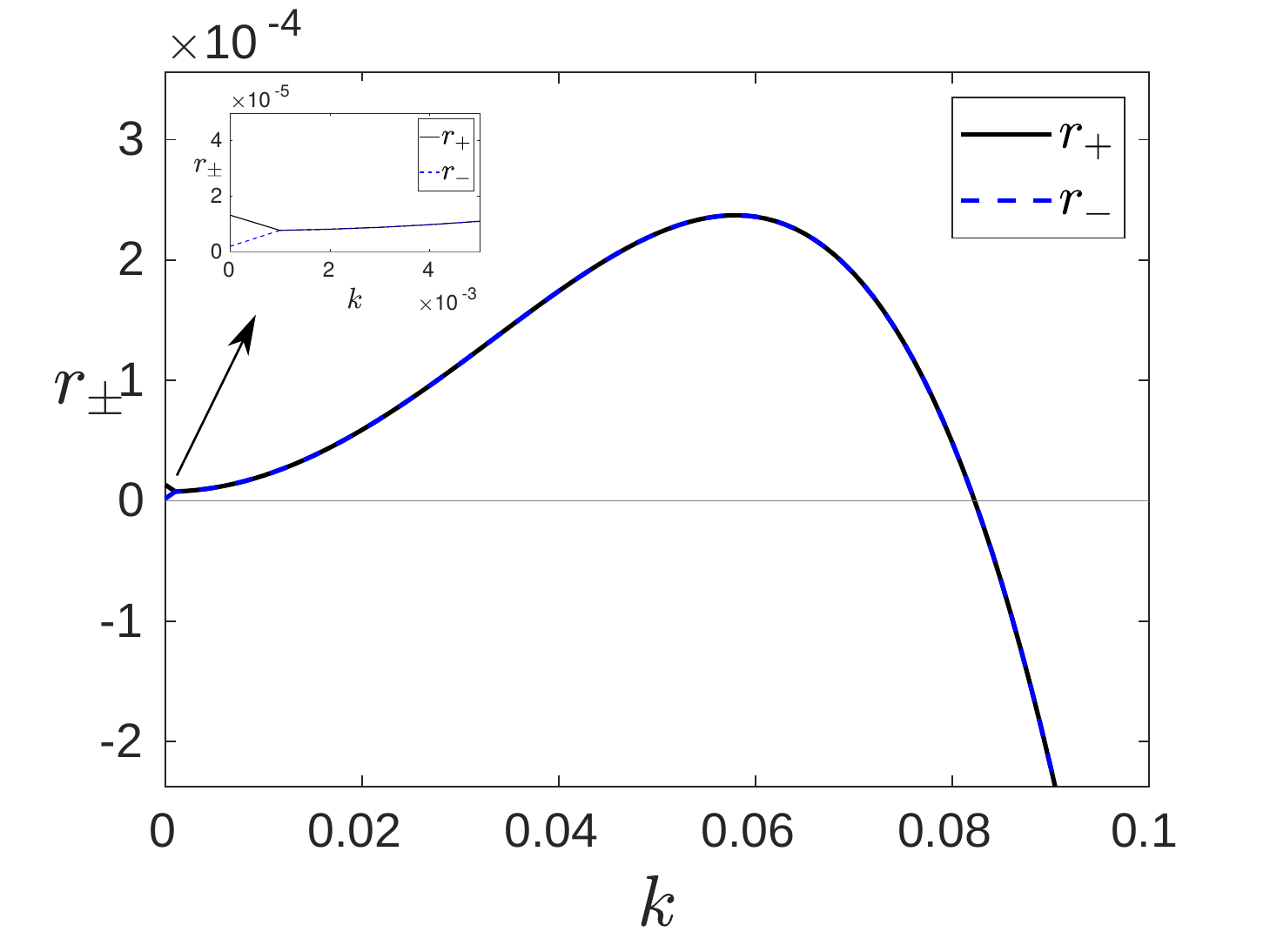}}
	\caption{\small The growth rates $r_+$ and $r_-$ versus the wavenumber k, for $E = 10^{-5}$. The remaining parameters are shown in Table \ref{tab:parameters1}.}
	\label{fig:section6A_growth_rate}
\end{figure}

With evaporation, the system has real eigenvalues for very small wavenumbers and complex eigenvalues for moderate wavenumbers, as shown in Fig. \ref{fig:section6A_growth_rate}. Here, the thermal Marangoni effect destabilises the liquid layer for small wavenumbers while the surface tension stabilises the liquid layer for high wavenumbers. In the discriminant $d_1$ we can see that the evaporation will lead the system to a monotonic instability at $k = 0$ while for small wavenumbers there is a competition between the thermal and solutal Marangoni numbers. If the thermal Marangoni effect dominates, the system will go through a monotonic instability while if the solutal Marangoni number dominates the instability will be oscillatory. In this case, the growth rate of the oscillatory mode is much higher than the evaporation rate and therefore we expect that the oscillatory mode will not decay as in the cases described by Overdiep \cite{Overdiep1986}, Howison et al. \cite{Howison1997} and Eres et al. \cite{Eres1999}.

\subsection{Parametric analysis}

A parametric analysis on the stability of the evaporating thin liquid layer is performed for the case of non-equilibrium evaporation ($K \neq 0$) in the limit of small evaporation number, $E = 10^{-5}$ . In this limit, the main mechanisms of instability during the evaporation are the thermocapillarity (thermal Marangoni effect) and the solutocapillarity (solutal Marangoni effect). Therefore, the effects of the thermal Marangoni number and the solutal Marangoni number as well as the volatility of the components on the instabilities are analysed. 

\begin{figure}[!t]
	\centerline{\includegraphics[width=8cm]{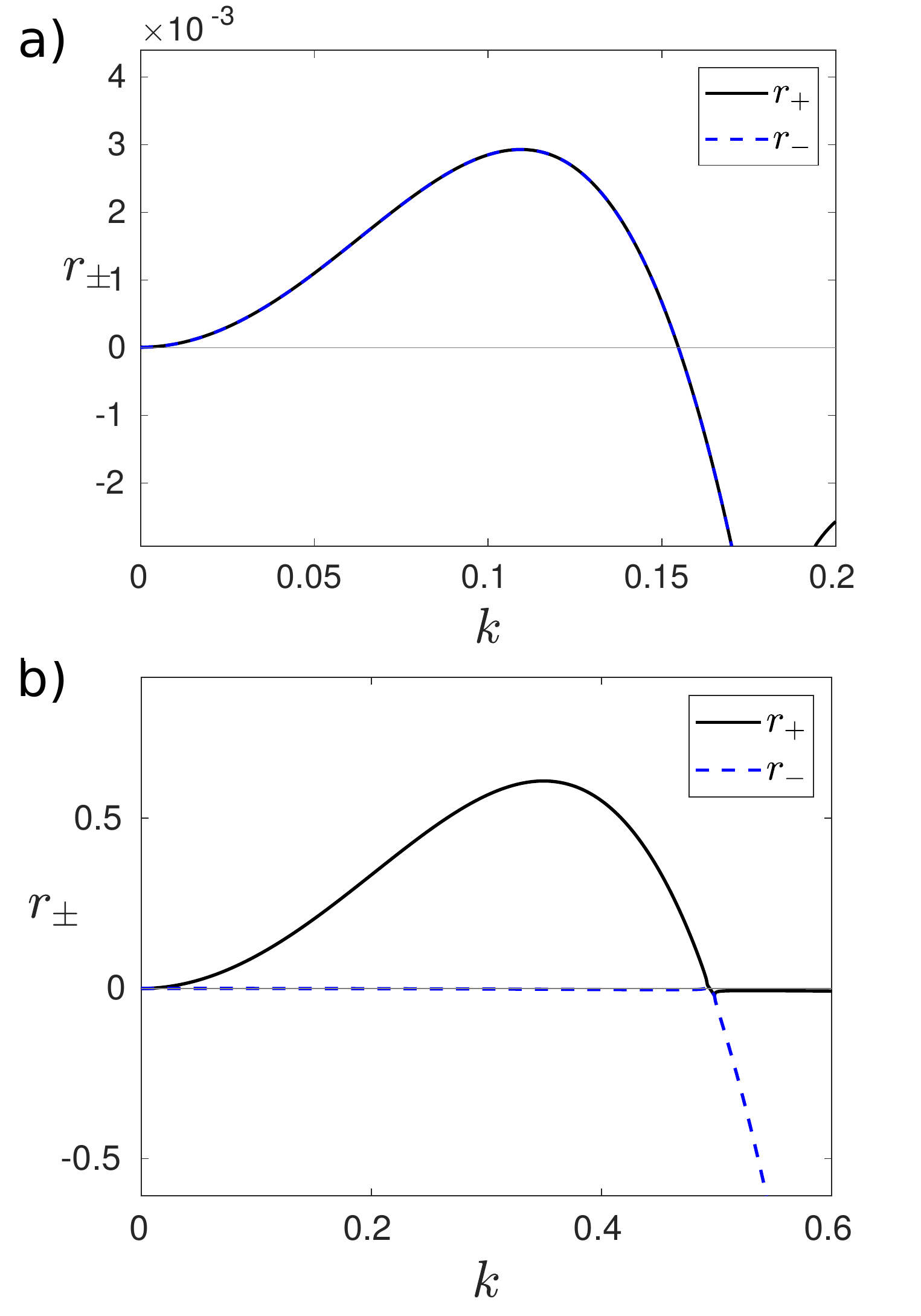}}
	\caption{\small Dependence on $M_T$: Growth rates $r_+$ and $r_-$ versus the wavenumber $k$ for $K \neq 0$ and $E = 10^{-5}$. Oscillatory instability mode for a) $M_T = 100$. Monotonic instability mode for b) $M_T = 1000$. The remaining parameters are shown in Table \ref{tab:parameters1}.}
	\label{fig:section6D_growth_rate_MT}
\end{figure}

Fig. \ref{fig:section6D_growth_rate_MT} presents the effect of the thermal Marangoni number on the instability. For small thermal Marangoni numbers the solutal Marangoni effect dominates reversing the initial perturbation and the evaporation process goes through an oscillatory instability mode, as seen in Fig. \ref{fig:section6D_growth_rate_MT}a. However, for high thermal Marangoni numbers the solutal Marangoni effect is not strong enough to reverse the initial perturbation. In this case the thermal Marangoni effect dominates promoting the initial perturbation and the evaporation process goes through a monotonic instability mode, as shown in Fig. \ref{fig:section6D_growth_rate_MT}b. Moreover, instabilities with shorter lengthscales are observed. It can be seen in Fig. \ref{fig:section6D_growth_rate_MT} that as the thermal Marangoni number increases both the growth rate and the wavenumber of the most unstable mode.

\begin{figure}[]
	\centerline{\includegraphics[width=8cm]{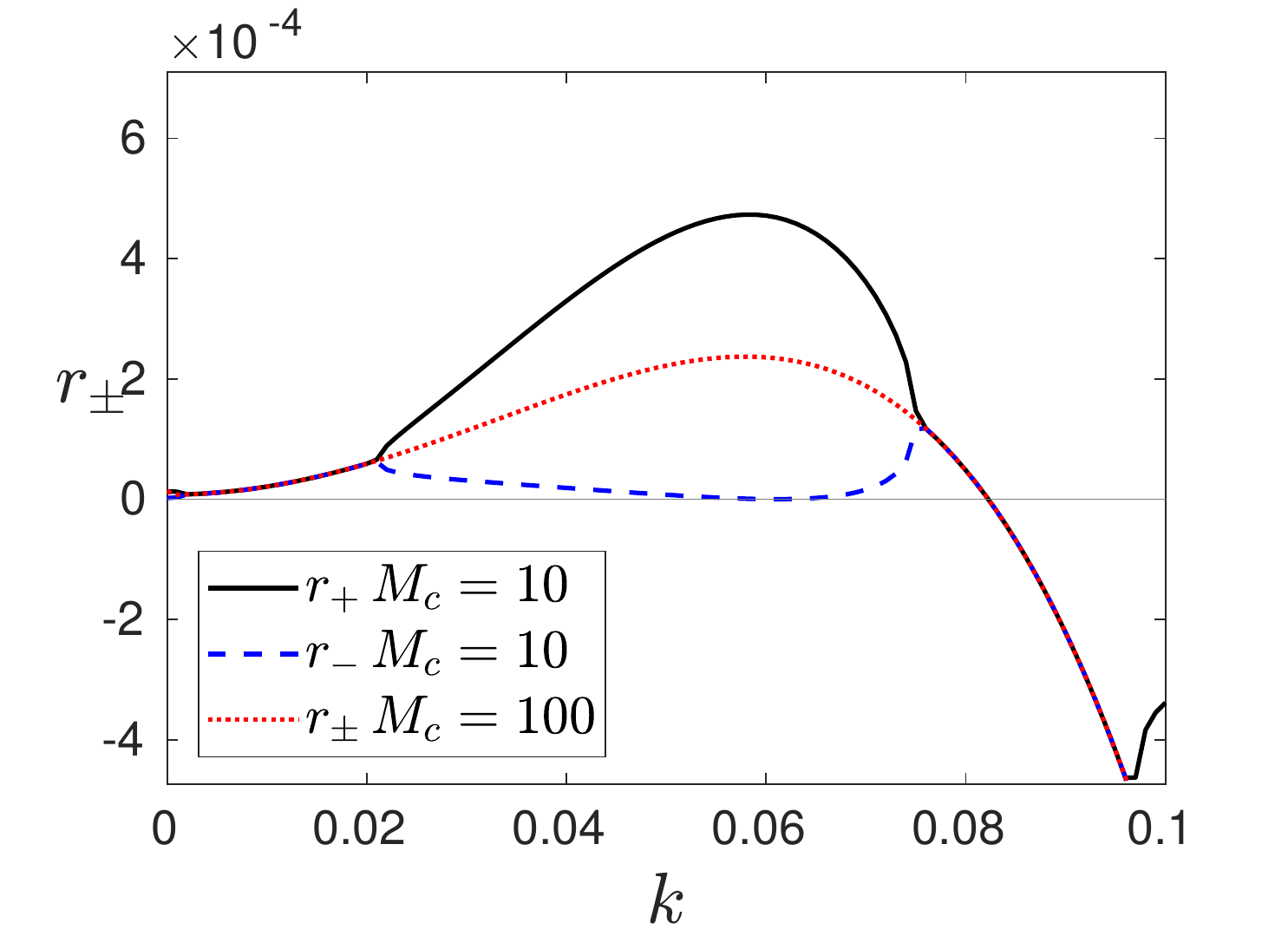}}
	\caption{\small Dependence on $M_c$: Growth rates $r_+$ and $r_-$ versus the wavenumber $k$ for $K \neq 0$ and $E = 10^{-5}$. Monotonic instability mode for $M_c = 10$ and oscillatory instability mode for $M_c = 100$. The remaining parameters are shown in Table \ref{tab:parameters1}.}
	\label{fig:section6D_growth_rate_Mc}
\end{figure}

The effect of the solutal Marangoni number on the instability is presented in Fig. \ref{fig:section6D_growth_rate_Mc}. For lower solutal Marangoni numbers, the thermal Marangoni effect dominates promoting the initial perturbation and the evaporation process goes through a monotonic instability mode for the most unstable wavenumber while the oscillatory mode is also unstable for short- and long-wave disturbances. For higher solutal Marangoni numbers, the solutocapillarity dominates over the thermocapillarity reversing the initial perturbation, and the evaporation process undergoes an oscillatory instability mode. It can be seen in Fig. \ref{fig:section6D_growth_rate_Mc} that the growth rate and wavenumbers are of the same order of magnitude for both cases. This shows that growth rate and the wavenumber of the most unstable mode is a stronger function of thermal Marangoni number (as seen from Fig. \ref{fig:section6D_growth_rate_MT}). We can also see that the growth rate of the monotonic instability is higher than the oscillatory instability for the most unstable wavenumber.

\begin{figure}[]
	\centerline{\includegraphics[width=8cm]{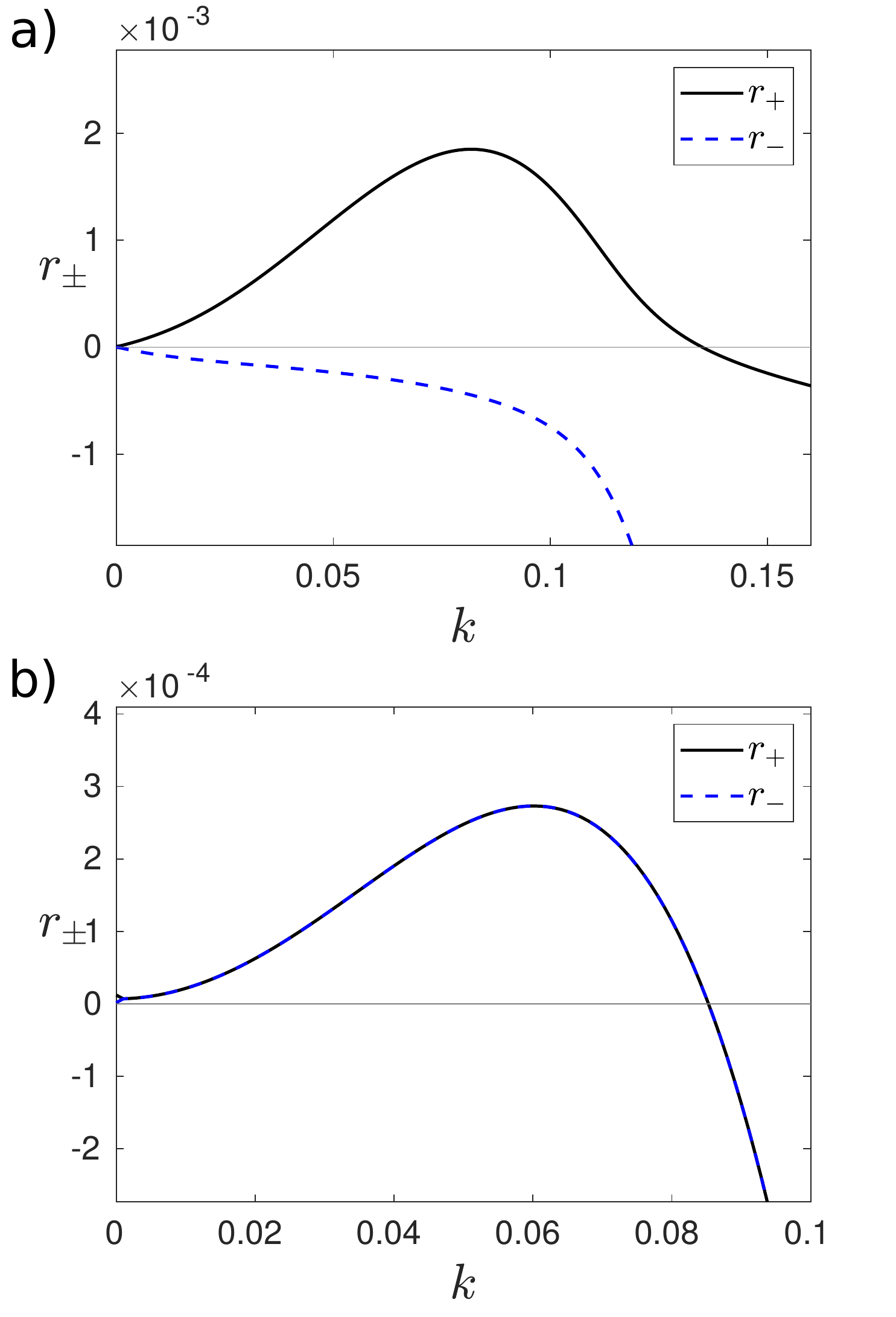}}
	\caption{\small Dependence on volatility: Growth rates $r_+$ and $r_-$ versus the wavenumber k for $K \neq 0$  and $E = 10^{-5}$. Monotonic instability mode for a) $\alpha = 0.5$. Oscillatory instability mode for b) $\alpha = 2$. The remaining parameters are shown in Table \ref{tab:parameters1}.}
	\label{fig:section6D_growth_rate_alpha}
\end{figure}

\begin{figure}[]
	\centerline{\includegraphics[width=8cm]{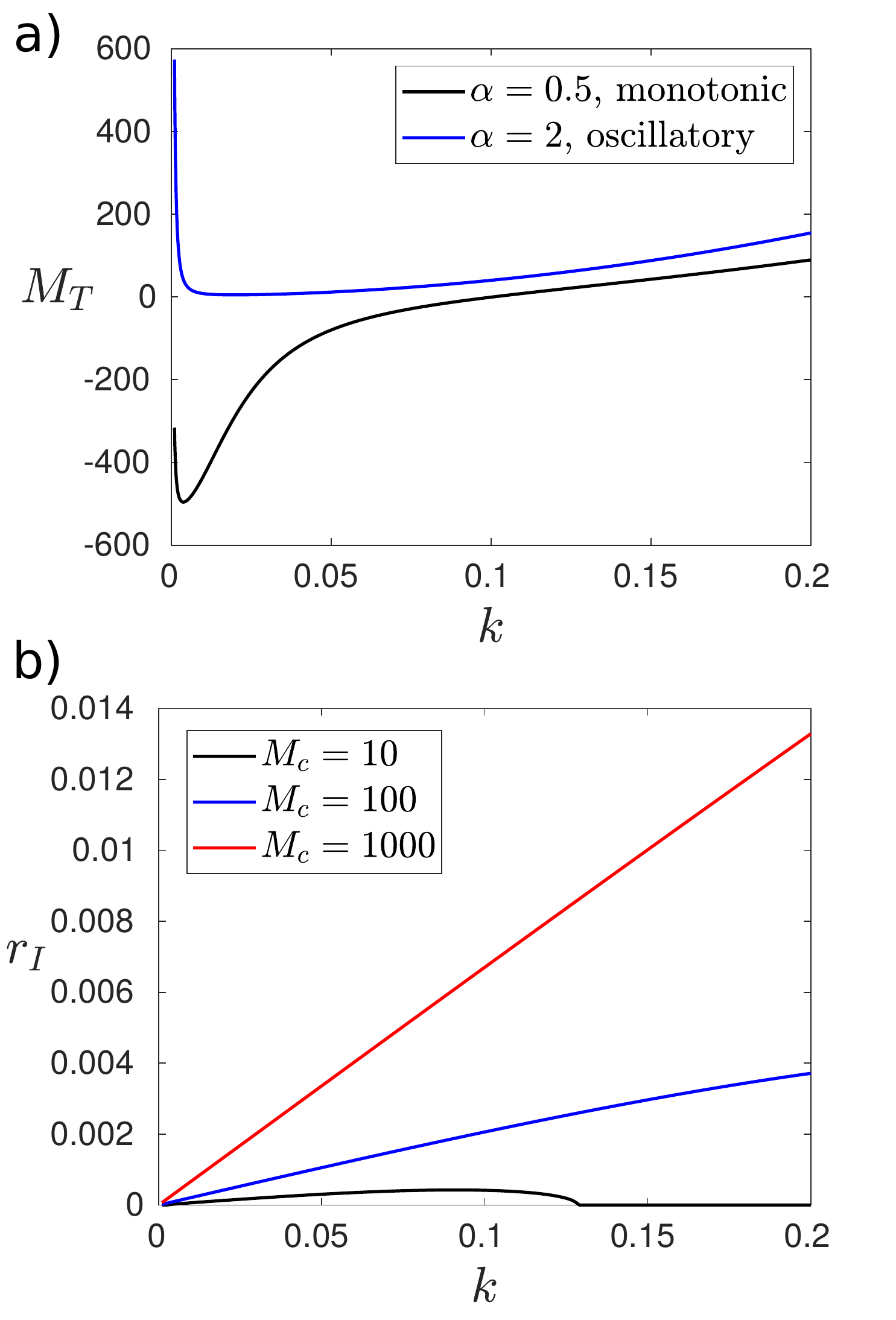}}
	\caption{\small a) Neutral curves for two different values of relative volatility leading to either a monotonic or an oscillatory mode of instability. b) Imaginary part of the most unstable eigenvalue for the oscillatory mode.}
	\label{fig:section6D_neutral_curve}
\end{figure}

Fig. \ref{fig:section6D_growth_rate_alpha} presents the effect of volatility of the components on the instability. For $\alpha = 0.5$ the volatility of component A is higher, so it evaporates faster at the trough increasing the volume fraction of component B. As the component B has lower surface tension ($\delta = 0.276$) the surface tension at the trough is reduced and the solutal Marangoni effect drives the liquid away from the trough in the direction of the crest that has higher volume fraction of component A, assisting the thermal Marangoni effect to promote the initial perturbation. In this case the evaporation process goes through a monotonic instability mode as shown in Fig. \ref{fig:section6D_growth_rate_alpha}a. For $\alpha = 2$, where component B is the most volatile, this component will evaporate first at the trough, increasing the volume fraction of component A. This will increase the surface tension at the trough and the solutal Marangoni effect will reverse the initial perturbation leading to an oscillatory instability mode, Fig. \ref{fig:section6D_growth_rate_alpha}b. It can be seen from Fig. \ref{fig:section6D_growth_rate_alpha} that as the relative volatility increases the growth rate of the most unstable wavenumber decreases as well as the most unstable wavenumber.

To determine the critical conditions for each mode of instability, we also present in Fig. \ref{fig:section6D_neutral_curve} the neutral curves for the same two different values of relative volatility. At this point, it is instructive to recollect that our quasi-steady state assumption considers that disturbances should have a much larger growth rate in comparison to evaporation rate. Thus, to derive our expressions for neutral stability, we assume that critical conditions arise when the real part of the eigenvalue is at least equal to the evaporation number $E$. The analytical expressions are presented in the Appendix B. Fig. \ref{fig:section6D_neutral_curve}a shows the neutral curves for the critical value of $M_T$ as a function of the wavenumber, Eq. (\ref{eq:neutral_mono}), while keeping the rest of the parameters constant. For $\alpha=0.5$ (monotonic case), it is shown that at small values of the wavenumber critical $M_T$ becomes negative; here, the solutal Marangoni number is $M_c=788$. Clearly, here, solutal gradients are able to destabilise the flow even without the presence of a thermal gradient. In fact, it is shown that in order to stabilise the flow the thermal gradient should be reversed, i.e. corresponding to negative values of $M_T$. At increasing values of the wavenumber, the lengthscale of the disturbance increases and therefore diffusion is able to smoothen out the solutal gradients which results in significant increase of the critical $M_T$. Turning our attention to $\alpha=2$, which corresponds to an oscillatory mode of instability, we note that according to Eq. (\ref{eq:neutral_osc}) the critical value of $M_T$ does not depend on $M_c$. However, the value of $M_c$ affects significantly the imaginary part of the most unstable eigenvalue and therefore, the frequency of the instability as depicted in Fig. \ref{fig:section6D_neutral_curve}b. Increasing $M_C$ leads to increase in the frequency of the instability.  

\subsection{Flow maps}

\begin{figure}[]
	\centerline{\includegraphics[width=8cm]{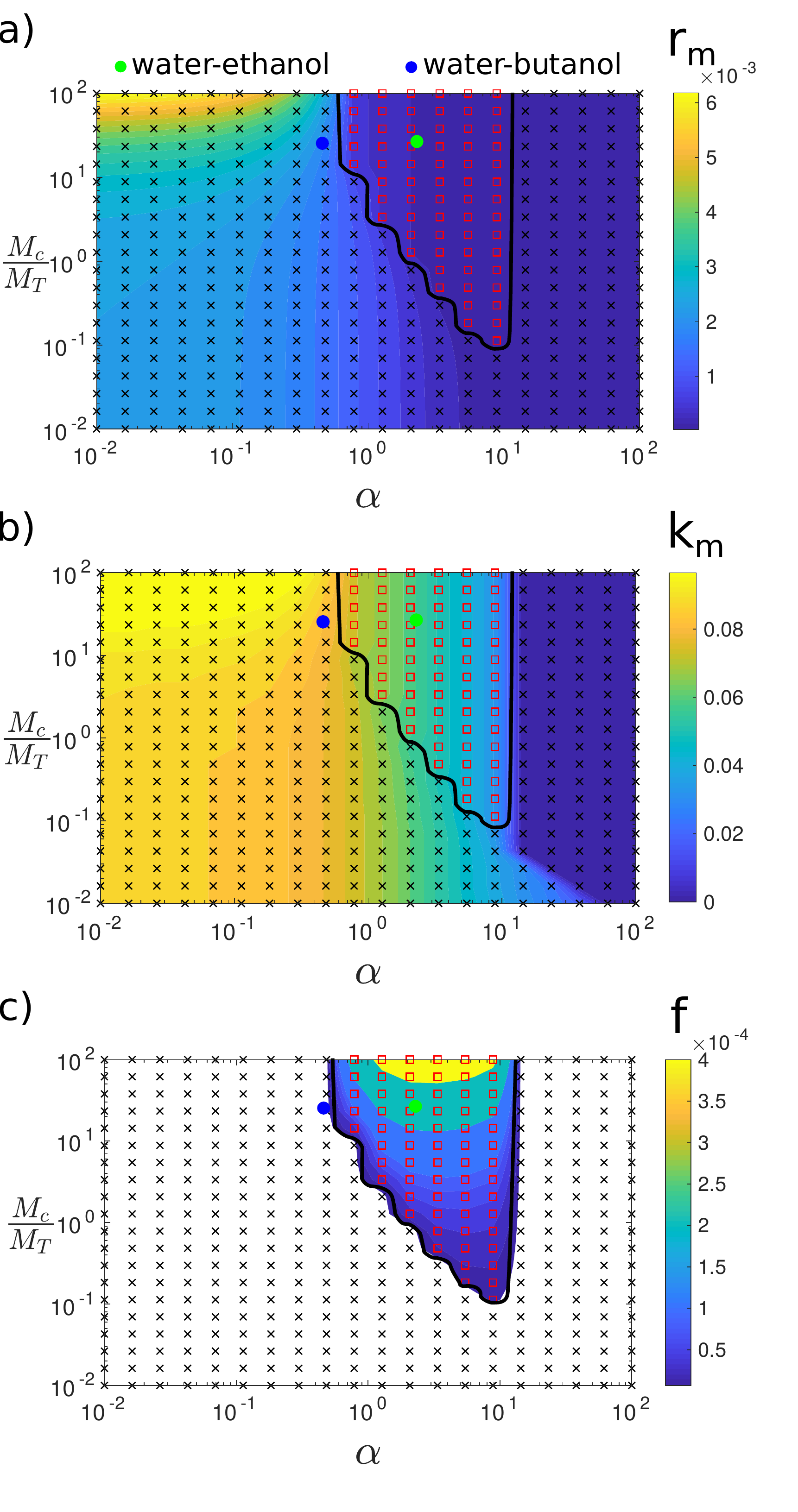}}
	\caption{\small Maps showing the regions of monotonic instability mode (crosses) and oscillatory instability mode (squares) in the parameter space of $M_c/M_T$ versus $\alpha$. a) Contour lines of the growth rate of the most unstable wavenumber. b) Contour lines of the most unstable wavenumbe. c) Contour lines of the frequency of the oscillations. Here $E = 10^{-5}$ and the remaining parameters are shown in Table \ref{tab:parameters1}. The experimental points for water-ethanol (parameters from Table \ref{tab:parameters1}) and water-butanol are plotted in green and blue bullets, respectively. The parameters used for butanol was $p^o = 22 \times 10^3$ and $\sigma = 19.45 \times 10^{-3}$.}
	\label{fig:section6D_map_E1e-5}
\end{figure}

Fig. \ref{fig:section6D_map_E1e-5} presents maps with the regions of monotonic and oscillatory instability modes in the parameter space of the relative volatility, $\alpha$, and the ratio of solutal and thermal Marangoni numbers, $M_c/M_T$. We choose the ratio between the solutal and thermal Marangoni numbers in order to compare solutocapillarity with thermocapillarity. The contours in Figs. \ref{fig:section6D_map_E1e-5}a, b and c show the growth rate, $r_m$, wavenumber, $k_m$, and the frequency, $f$, respectively of the most unstable mode. 

In Fig. \ref{fig:section6D_map_E1e-5} the component $A$ has higher surface tension ($\delta = 0.27$), therefore solutocapillarity drives the liquid in direction to regions with higher volume fraction of component A. It can be seen in Fig. \ref{fig:section6D_map_E1e-5} that for the case where component A is less volatile, $\alpha > 1$, a smaller ratio between the solutal and thermal Marangoni numbers is needed to achieve the oscillatory instability mode as $\alpha$ increases. This is because the solutocapillarity is proportional to how fast the lower surface tension component evaporates compared to the one with higher surface tension. Therefore increasing relative volatility requires a lower ratio between the solutal and thermal Marangoni number for solutocapillarity to overcome the thermocapillarity leading the system to an oscillatory instability mode. For $\alpha > 10$ the oscillatory mode is overhauled by the long-wave monotonic mode due to evaporation.

From Figs. \ref{fig:section6D_map_E1e-5}a and b it can be seen that the growth rate and the wavenumber of the most unstable mode decrease as the relative volatility increases and for $\alpha < 1$ they increase as the ratio of the solutal and thermal Marangoni numbers increases. In Fig. \ref{fig:section6D_map_E1e-5}b we can see that for $\alpha > 10$ the most unstable instability become monotonic and $k_m = 0$ because the growth rate of the oscillatory instability decreases with $\alpha$ as show in Fig. \ref{fig:section6D_growth_rate_alpha}. Fig. \ref{fig:section6D_map_E1e-5}c shows that the frequency of the oscillations increases with the solutal and thermal Marangoni numbers ratio and has a maximum around $\alpha \approx 3$.

\subsection{Mechanisms of the instability}

\begin{figure}[]
	\centerline{\includegraphics[width=8cm]{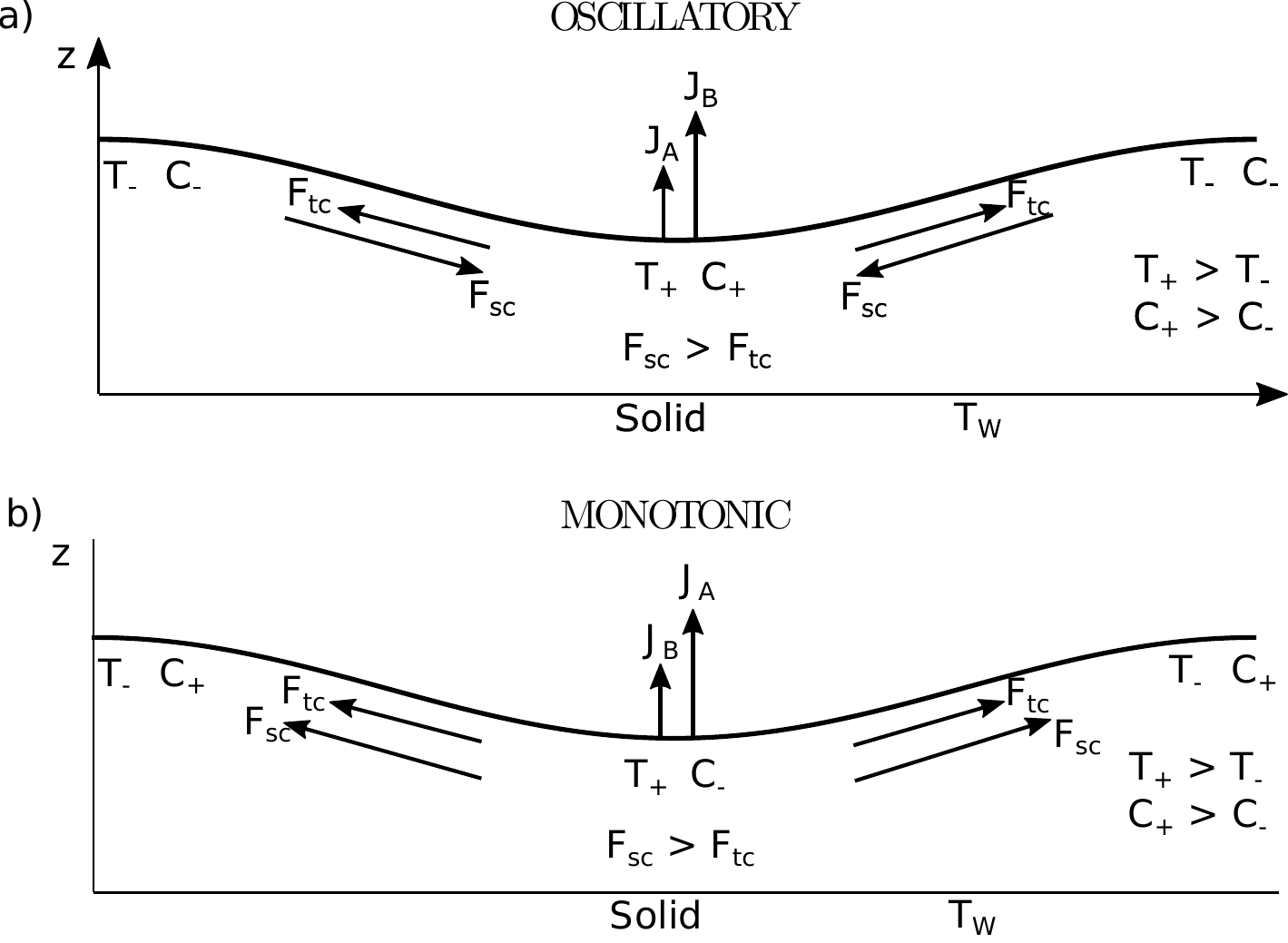}}
	\caption{\small a) Schematic of the oscillatory instability for the case where component A is less volatile than B and has higher surface tension. b) Schematic of the monotonic instability for the case where component A is more volatile than B and has higher surface tension. Here $F_{tc}$ is the thermocapillary force and $F_{sc}$ is the solutocapillary force.}
	\label{fig:section6C_sketch_instability}
\end{figure}

A schematic of the main mechanisms of instability is shown in Fig. \ref{fig:section6C_sketch_instability}. First we discuss the instability for the case of the standard parameters present in Table \ref{tab:parameters1} where the component A has lower volatility and higher surface tension than component B. When an initial perturbation is applied to the system the temperature of the interface becomes hotter at the trough due to the proximity to the hot substrate. Therefore, the thermal Marangoni effect drives the liquid from the hotter trough in direction to the colder crest promoting the perturbation. However, due to the higher volatility of component B, it evaporates faster at the trough increasing the volume fraction of component A that has higher surface tension. As the volume fraction of component A increases at the trough, the solutal Marangoni effect becomes stronger and at some point it may overcome the thermal Marangoni effect and starts to drive the liquid in the direction of the trough, as shown in Fig. \ref{fig:section6C_sketch_instability}a. As a consequence, the interface starts to level until the trough become a crest and the previous crests become troughs. This process repeats at the new troughs causing oscillations at the interface and the evaporation goes through an oscillatory instability mode. However, when component A has higher volatility and higher surface tension than component B the solutal Marangoni effect has the opposite behaviour. The volume fraction of component B increases at the trough due to the higher volatility of component A and the solutal Marangoni effect drives the liquid from the trough in direction to the crest, promoting the thermal Marangoni effect, as shown in Fig. \ref{fig:section6C_sketch_instability}b. In this case the evaporation goes through a monotonic instability mode. Thus, when there is a competition between thermal and solutal Marangoni effects the oscillatory instability is possible only when solutocapillarity overcomes thermocapillarity. On the other hand, when the thermal and solutal Marangoni effects enhance each other driving the flow in the same direction the instability is always monotonic. 

\begin{figure}[]
	\centerline{\includegraphics[width=7.5cm]{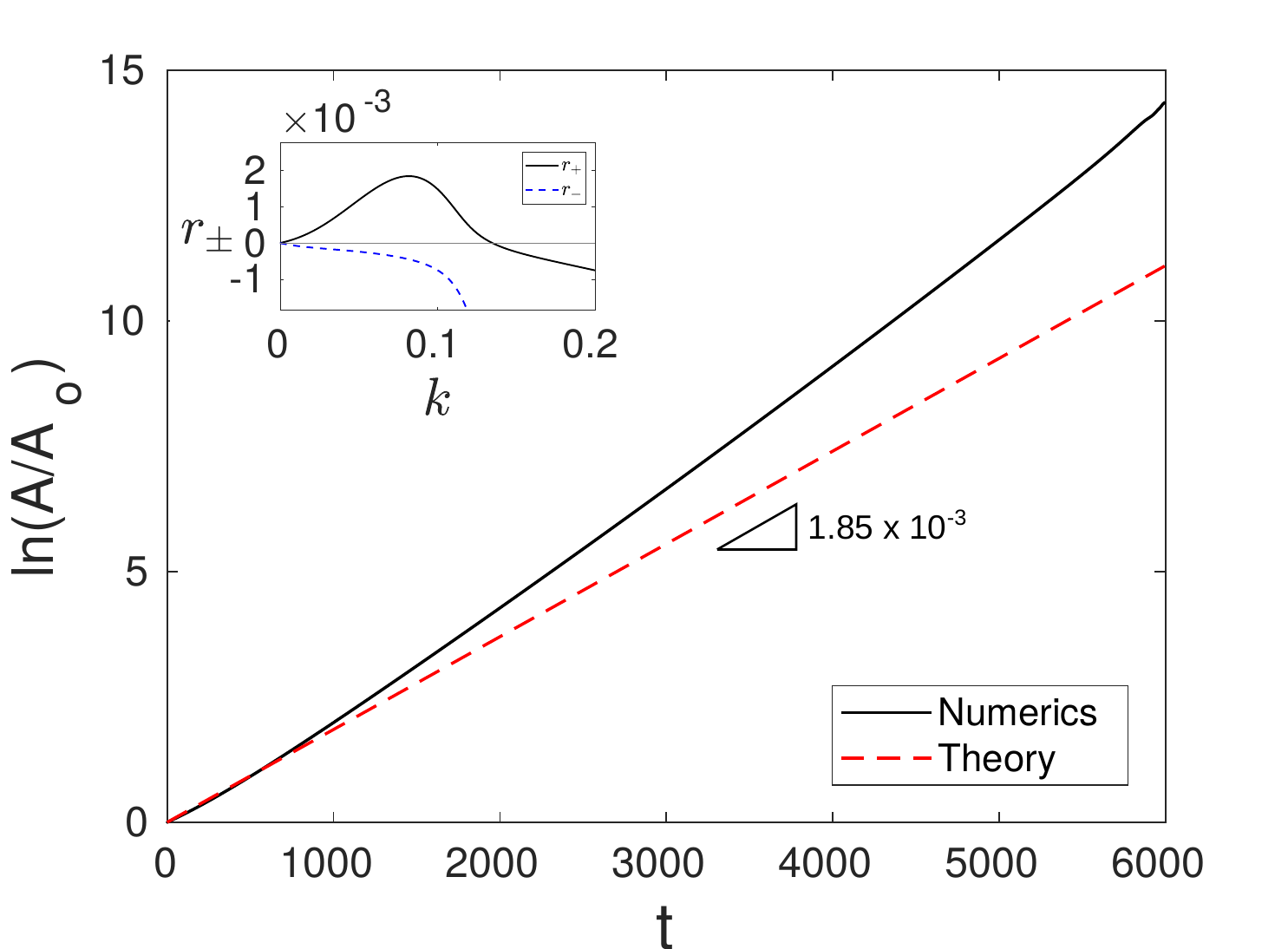}}
	\caption{\small Growth rate of the amplitude of the initial perturbation over time derived from the transient simulation. Inset with the growth rate versus the wavenumber derived from the linear stability analysis for the case of monotonic instability mode with $\alpha = 0.5$. Here $E = 10^{-5}$ and the remaining parameters are shown in Table \ref{tab:parameters1}.}
	\label{fig:section7_validation_mon}
\end{figure}

\section{Non-Linear regime}

\subsection{Validation against linear theory}

We examine the non-linear dynamics by solving the evolution Eqs. (\ref{eq:LW_h_t2}) and (\ref{eq:LW_C_t2}), dynamically. The dynamic solution so generated represent transient simulations of the problem. Cross-validation against linear stability analysis is presented in Section VII.B. The growth rates derived from the linear stability analysis and from the transient simulations are compared for the cases of monotonic instability mode and oscillatory instability mode. The growth rate of the initial perturbation in the transient simulations is calculated using the following expression for the monotonic and oscillatory case, respectively,
\begin{align}
&r_{mon} = \frac{\dot{A}}{A} \\
&r_{osc} = \frac{\dot{A}_{max}}{A_{max}}
\end{align}
Here, in the monotonic case, $A$ is the amplitude of the instability and $\dot{A}$ is its time derivative, and, in the oscillatory case, $A_{max}$ is the maximum amplitude of the oscillations and $\dot{A}_{max}$ is its time derivative. 

\begin{figure}[]
	\centerline{\includegraphics[width=8cm]{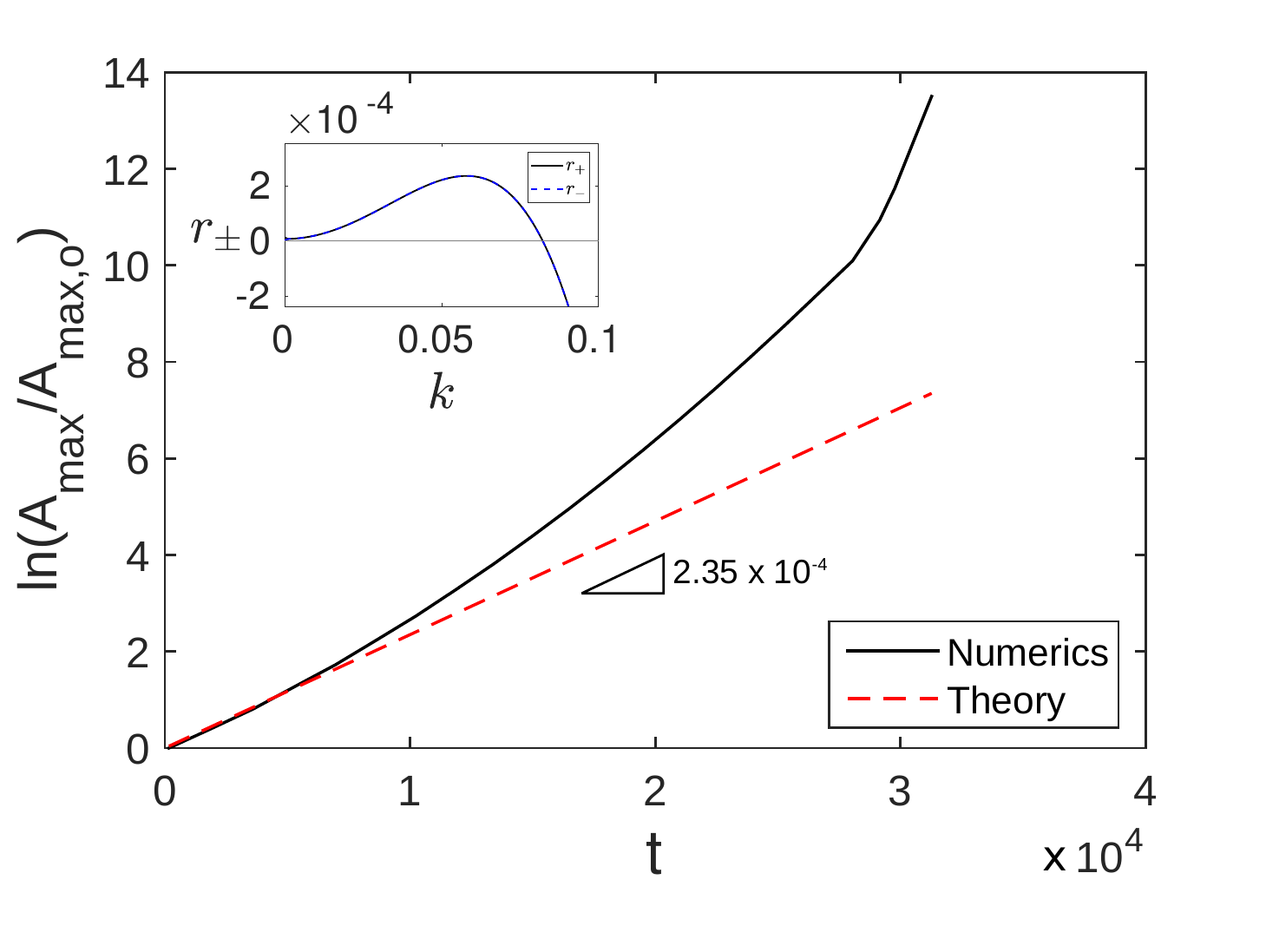}}
	\caption{\small Growth rate of the maximum amplitude of the oscillations over time derived from the transient simulation. Inset with the growth rate versus the wavenumber derived from the linear stability analysis for the case of oscillatory instability mode with $\alpha = 2.28$. Here $E = 10^{-5}$ and the remaining parameters are shown in Table \ref{tab:parameters1}.}
	\label{fig:section7_validation_osc}
\end{figure}

The evolution of the system is solved numerically with a bespoke code using the finite element method and the weak formulation of the equations. The computational domain is discretized in space using 100 elements and the solution is advanced in time using the implicit Euler method. The resulting set of nonlinear algebraic equations are solved in each time step using the Newton-Raphson method. Convergence was achieved upon mesh refinement. The size of the domain comprises the interval $0 < X < \pi/k_M$, where $k_M$ is the most unstable wave number. Periodic boundary conditions are applied on the lateral endings and we use the following initial condition, 
\begin{align}
h(X,0) = 1 + 5\times10^{-7}Cos(k_MX) \\
C(X,0) = 0.5 + 5\times10^{-7}Cos(k_MX)
\end{align}

First, we consider a case which exhibits an monotonic instability. In the inset of Fig. \ref{fig:section7_validation_mon} we depict the dispersion curve where it is shown that the most unstable wavenumber is at $k =  0.082$ corresponding to a growth rate of $r = 1.85 \times 10^{-3}$. We perform a transient simulation for a domain with size that is equal to the wavelength of the most unstable mode and evaluate the growth rate. The transient simulation shows a good agreement in growth rate for early times when the linear regime is still valid, as shown in Fig. \ref{fig:section7_validation_mon}. At later times, when waves become non-sinusoidal the numerical growth rate demonstrates a strong non-linear growth. 

In Fig. \ref{fig:section7_validation_osc} we compare the growth rates for the oscillatory instability mode. From linear stability analysis, inset of Fig. \ref{fig:section7_validation_osc}, the most unstable mode at $k = 0.058$ corresponds to a growth rate of $r = 2.35 \times 10^{-4}$. Again, as seen in the case for monotonic instability above, Fig. \ref{fig:section7_validation_osc} demonstrates that the growth rate predicted by the transient simulation agrees with that predicted from the linear stability analysis at early times. Both Fig. \ref{fig:section7_validation_osc} and Fig. \ref{fig:section7_validation_mon}, also show that the non-linear behaviour begins approximately when the perturbation amplitude is approximately an order of magnitude higher the initial value.

\subsection{Numerical simulations}

\begin{figure}[]
	\centerline{\includegraphics[width=8cm]{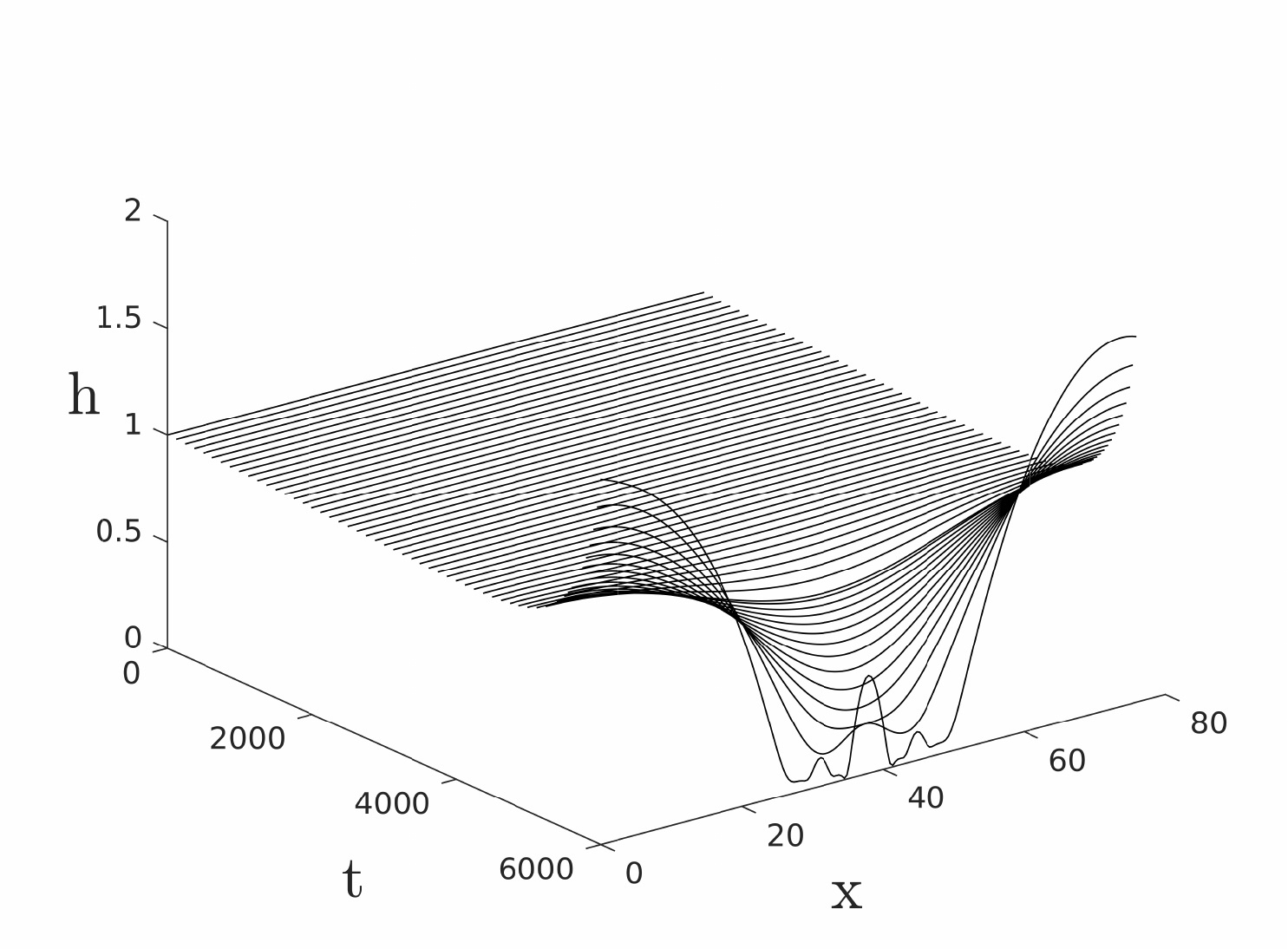}}
	\caption{\small Time evolution of the interface for $\alpha = 0.5$ and $E = 10^{-5}$ showing the monotonic instability mode. The remaining parameters are shown in Table \ref{tab:parameters1}.}
	\label{fig:section8_interface_evo_mon}
\end{figure}

The time evolution of the interface of a thin liquid layer composed of a binary mixture heated from below using the set of base parameters presented in Table \ref{tab:parameters1}, $E = 10^{-5}$ and $\alpha = 0.5$ (Fig. \ref{fig:section7_validation_mon}) is presented in Fig. \ref{fig:section8_interface_evo_mon}. For this set of parameters the component A has higher volatility and higher surface tension than component B. Initially a small perturbation, $O(10^{-6})$, is applied and it grows exponentially as the interface evaporates. When the perturbation is applied the temperature of the interface at the trough becomes hotter. The perturbation is then promoted by the thermocapillarity that drives the liquid away from the hotter trough, and by solutocapillarity due to the higher evaporation rate of component A at the trough which decreases the surface tension at that location. In this case, the evaporation process presents a monotonic instability. The time taken for the rupture of the liquid layer was $t_R = 5.99 \times 10^{3}$.

\begin{figure}[]
	\centerline{\includegraphics[width=8cm]{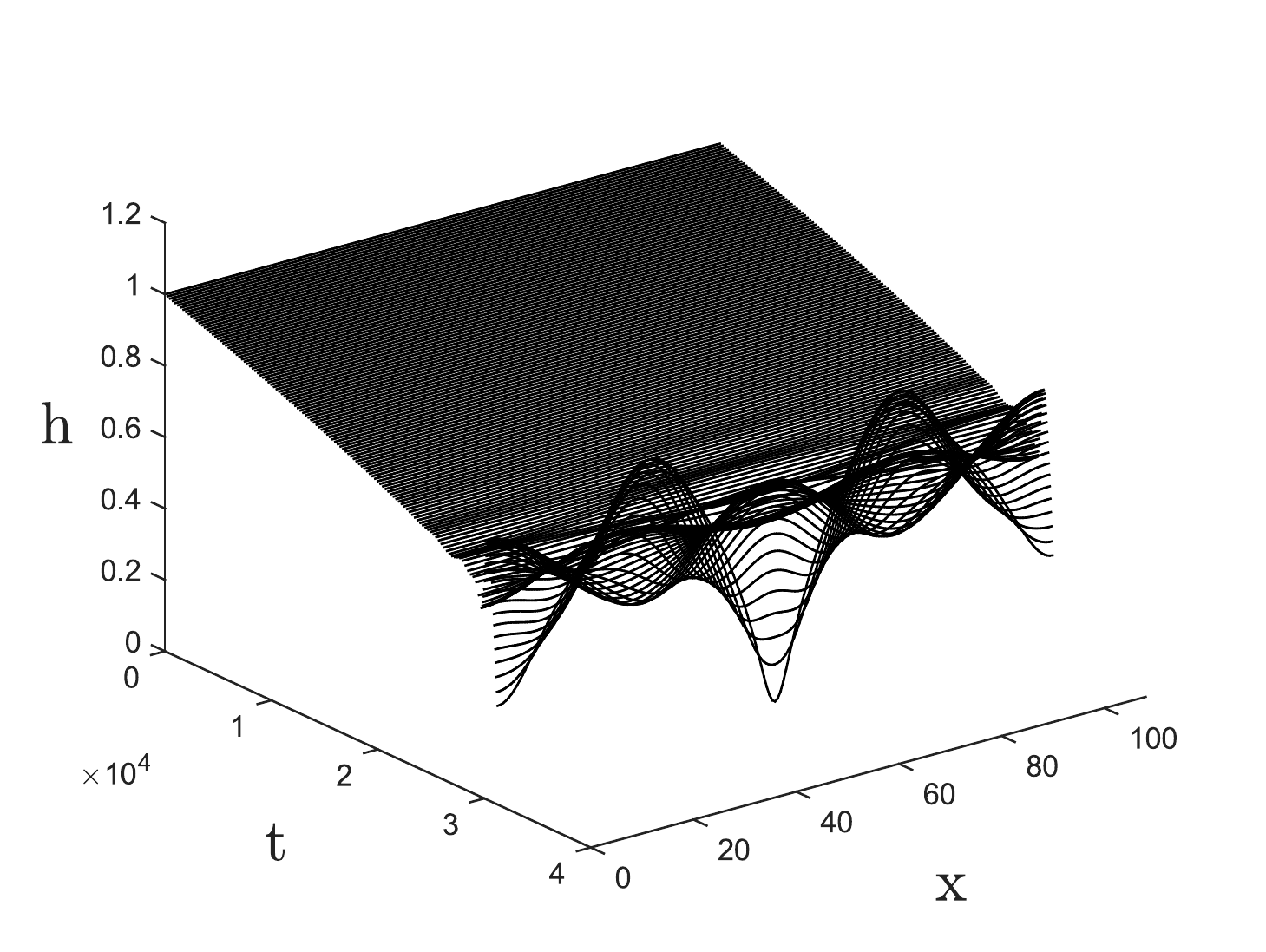}}
	\caption{\small Time evolution of the interface for $\alpha = 2.28$ and $E = 10^{-5}$ showing the oscillatory instability mode. The remaining parameters are shown in Table \ref{tab:parameters1} }
	\label{fig:section8_interface_evo_osc}
\end{figure}


Fig. \ref{fig:section8_interface_evo_osc} presents the time evolution of the interface for the set of base parameters present in Table \ref{tab:parameters1} and $E = 10^{-5}$ that corresponds to the oscillatory instability mode (Fig. \ref{fig:section7_validation_osc}). For this set of parameters the component A has lower volatility and higher surface tension than component B. In this case, the initial perturbation, $O(10^{-6})$, is initially promoted by the thermal Marangoni effect due to the higher temperature at the trough. However, as the volume fraction of component A increases at the trough due to the higher volatility of component B, the solutocapillarity becomes stronger and starts to drive the liquid in direction to the trough, reversing the amplitude of the initial perturbation. This process repeats at the new trough causing oscillations at the interface. The computed rupture time was $t_R = 3.15 \times 10^{4}$. In order to test the effect of the size of the domain, we considered this case with a double size domain. The evolution of the interface presents the development of the same structures in both case showing no dependence on the size of the domain.

\begin{figure}[]
	\centerline{\includegraphics[width=8cm]{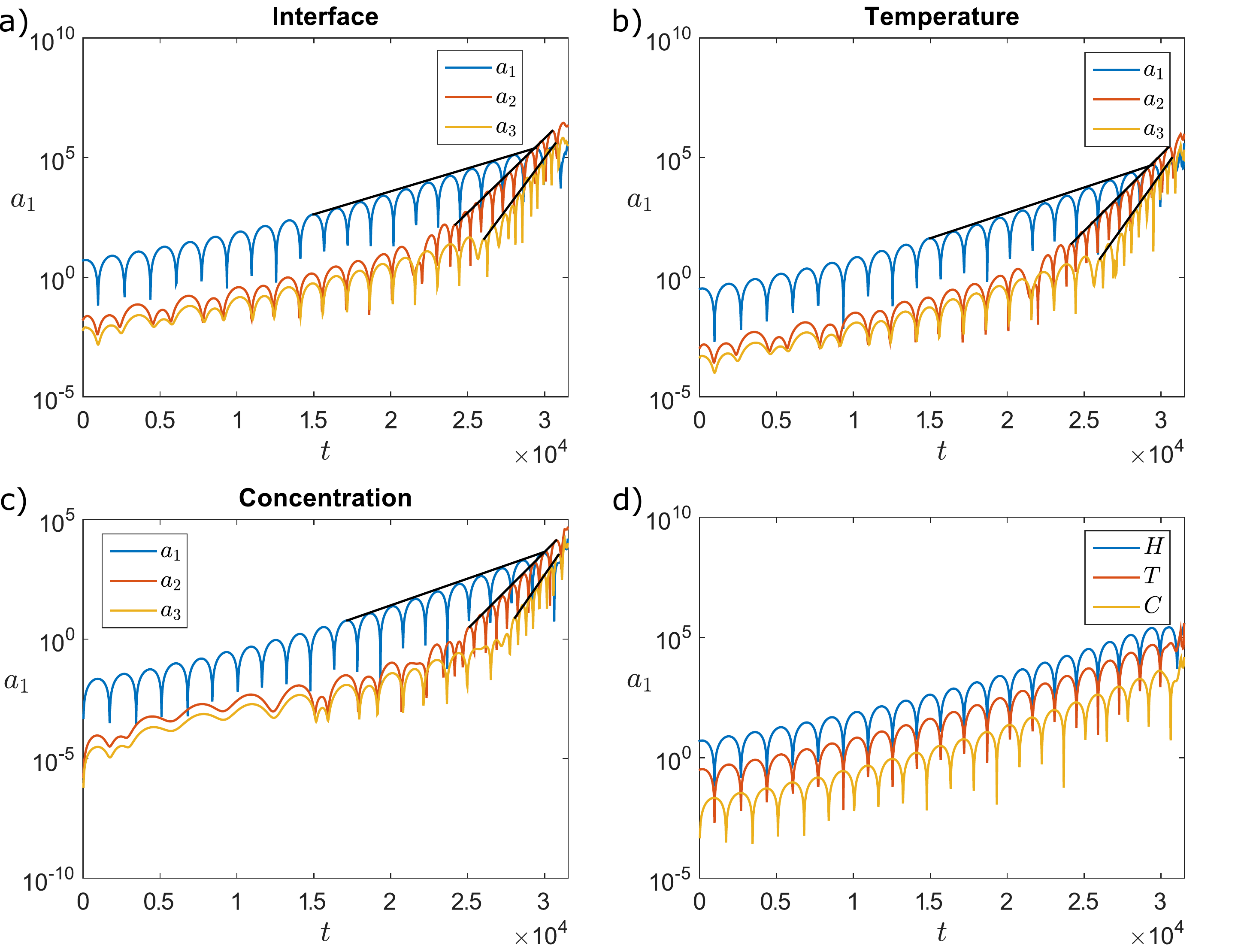}}
	\caption{\small Evolution of higher order Fourier modes for transient simulations presented for oscillatory instability in Fig. \ref{fig:section8_interface_evo_osc},  a) the interface ($s_1 = 2.21 \times 10^{-4}$, $s_2 = 7.20 \times 10^{-4}$, $s_3 = 9.92 \times 10^{-4}$), b) the temperature ($s_1 = 2.44 \times 10^{-4}$, $s_2 =  7.36 \times 10^{-4}$, $s_3 = 1.04 \times 10^{-3}$) and c) the volume fraction ($s_1 = 2.57 \times 10^{-4}$, $s_2 = 7.40 \times 10^{-4}$, $s_3 = 1.07 \times 10^{-3}$). d) First mode of the Fourier transform of the interface (H), temperature (T) and volume fraction (C).}
	\label{fig:section8_modes}
\end{figure}

Figs. \ref{fig:section8_modes}a, b, and c show the evolution of higher order modes obtained by the Fourier transform of the interfacial height, interfacial temperature and interfacial volume fraction predicted by the numerical simulations for the case with oscillatory instability presented in Fig. \ref{fig:section8_interface_evo_osc}a. Each subsequent mode that appears has a higher slope than the previous one. The ratio between the slopes of the modes for interface deformation are $s_2/s_1 = 3.23$ and $s_3/s_1 = 4.49$. For temperature these are $s_2/s_1 = 3.01$ and $s_3/s_1 = 4.24$, and for volume fraction these are $s_2/s_1 = 2.88$ and $s_3/s_1 = 4.16$. In Fig. \ref{fig:section8_modes}d it can be seen that the first or the primary modes of the interface and temperature are in phase, while that of volume fraction is out of phase. It means that the temperature changes instantaneously with the interface, increasing at the trough and decreasing at the crest. However, volume fraction is significantly out of phase. This means that as the interface is enriched by the less volatile component, evaporation ensures local cooling of the interface - thereby leading to conclude that the interfacial and thermal fluctuations are enslaved to the soluto-Marangoni instability.

\section{Conclusions}

The stability of the evaporation of an horizontal thin liquid layer comprising a binary mixture of volatile liquids heated from below has been investigated by means of linear stability analysis and transient simulations. To that effect, long-wave approximation has been employed to derive the evolution equations for the free interface and the volume fraction of the components. The linear stability analysis has been cross-validated by comparing the solution for the growth rate of the instabilities against the transient simulations.

Two modes of instabilities have been described by the linear theory, i.e. a monotonic instability mode and an oscillatory instability mode. By performing a parametric analysis it was possible to identify how these modes depend on the ratio between the thermal and solutal Marangoni number and on the relative volatility. When the most volatile component has the lower surface tension the thermal and solutal Marangoni effects compete with each other. In this case, when the solutal Marangoni effect dominates the system presents an oscillatory intability mode. However, when the thermal Marangoni effect dominates the system presents a monotonic instability mode. On the other hand, when the most volatile component has the higher surface tension both the thermal and the solutal Marangoni effects assist each other promoting the initial perturbation and leading to a monotonic instability mode.

\section{Acknowledgements}
The authors gratefully acknowledge the supports received from  ThermaSMART project of European Commission (Grant no. EC-H2020-RISE-ThermaSMART-778104). GK acknowledges the support received by the SPREAD project of Hellenic Foundation for Research and Innovation and General Secretariat for Research and Technology (Grant no. 792). 

\appendix
\section{Base State}
Here we derive the base state solution. Under the assumptions stated in Section IV the momentum, energy and volume fraction conservation equations become,
\begin{align}
\bar{p}_z &= 0 \\
Pr(c_p \bar{T})_t &= (\lambda \bar{T}_z)_z \\
\bar{c}_t &= \frac{\bar{c}_{zz}}{Pe}
\end{align}

At the interface $z = h(t)$ the energy and the normal stress balance become,
\begin{gather}
\bar{J}_A + \Lambda \bar{J}_B + \dfrac{E^2}{2 \mathcal{L}D^2} \bar{J}^3 = - \lambda \bar{T}_z \\
\bar{p} = p_v + \dfrac{ E^2 \bar{J}^2}{D} + \dfrac{\mathcal{A}}{\bar{h}^3}
\end{gather}

\noindent
there is no shear stress in the base state. The volume fraction boundary condition becomes,
\begin{equation}
\frac{\bar{c}_z|_{z=\bar{h}}}{Pe} = E (\bar{c} J - J_A)
\end{equation}

The kinematic boundary condition becomes,
\begin{equation}
E \bar{J} = - \bar{h}_t
\end{equation}

\noindent
while the constitutive equation for the evaporation flux is given by,
\begin{align}
K\bar{J}_A &= \bar{c} \bar{T} \\
K\bar{J}_B &= (1 - \bar{c})\alpha\beta^{\frac{3}{2}} \Lambda \bar{T} 
\end{align}

At the solid boundary $z = 0$, the boundary condition is,
\begin{equation}
\bar{T} = 1
\end{equation}

Since we consider a slowly evaporating film $E$ is considered to be small and to retain the effect of mass loss in the kinematic boundary condition, time is rescaled on the evaporative scale,
\begin{equation}
t' = Et, \qquad z' = z
\end{equation}

The total mass flux $\bar{J}(t')$ and the liquid temperature $\bar{T}(z',t')$ are considered to be of order unity, while pressure $\bar{p}(t')$ of order $E^{-1}$. These dependent variables are expanded in power of $E$,
\begin{align}
&\bar{c} = c_o + Ec_1 + E^2c_2 + ... \\
&\bar{J_A} = J_{Ao} + EJ_{A1} + E^2J_{A2} + ... \\
&\bar{J_B} = J_{Bo} + EJ_{B1} + E^2J_{B2} + ... \\
&\bar{T} = T_o + ET_1 + E^2T_2 + ... \\
&\bar{p} = E^{-1} ( p_o + Ep_1 + E^2p_2 + ... )
\end{align}

\noindent
while the film thickness $\bar{h}(t')$ is considered an unspecified order-one function.

We assume $\mathcal{L} \gg 1$ in order to neglect the kinetic energy in the energy balance. Lets assume $A = \bar{A}E^{-1}$ in order to keep the disjoining pressure in the normal-stress balance, where $\bar{A}$ is an order one quantity.

Applying the time rescaling and substituting the expansions on the base state, in the small $E$ limit the leading-order base state system becomes
\begin{align}
& \quad p_{o,z'} = 0 \\
& \quad (\lambda T_{o,z'})_{z'} = 0 \\
& \quad \bar{c}_{o,t'} = \frac{\bar{c}_{1,z'}|_{z'=\bar{h}}}{\bar{h}} \\
\text{At } z' = h(t') \quad: &\quad J_o = - \bar{h}_{t'} \\
& \quad J_{o,A} + \Lambda J_{o,B} = - \lambda T_{o,z'} \\
& \quad \frac{p_o}{E} = \frac{A}{\bar{h}^3} \\
& \quad KJ_{o,A} = \bar{c} T_o \\
& \quad KJ_{o,B} = (1 - \bar{c})\alpha\beta^{\frac{3}{2}} \Lambda T_o \\
& \quad \bar{c}_{1,z'}|_{z'=\bar{h}} = \bar{c}_o J_o - J_{A,o} \\
\text{At } z' = 0 \qquad: & \quad T_o = 1
\end{align}

\noindent
along with the initial condition
\begin{equation}
t'= 0, \qquad \bar{h} = 1
\end{equation}

The resulting leading-order base state solution is
\begin{gather}
\bar{h} = - \dfrac{\bar{\lambda} K}{\bar{\Lambda}_2}+ \dfrac{1}{\bar{\Lambda}_2}\sqrt{ (\bar{\lambda} K + \bar{\Lambda}_2)^2 - 2\bar{\lambda} \bar{\Lambda}_1\bar{\Lambda}_2 Et}
\label{eq:BS_h} \\
\bar{T} = 1 - \dfrac{\bar{\Lambda}_2 z}{\sqrt{ (\bar{\lambda} K + \bar{\Lambda}_2)^2 - 2\bar{\lambda} \bar{\Lambda}_1 \bar{\Lambda}_2 Et}}
\label{eq:BS_T} \\
\bar{J}_A = \dfrac{\bar{\lambda} \bar{c} }{\sqrt{ (\bar{\lambda} K + \bar{\Lambda}_2)^2 - 2\bar{\lambda} \bar{\Lambda}_1 \bar{\Lambda}_2 Et}}
\label{eq:BS_JA} \\
\bar{J}_B = \dfrac{\bar{\lambda} (1 - \bar{c})\alpha\beta^{\frac{3}{2}}\Lambda }{\sqrt{ (\bar{\lambda} K + \bar{\Lambda}_2)^2 - 2\bar{\lambda}  \bar{\Lambda}_1 \bar{\Lambda}_2 Et}}
\label{eq:BS_JB} \\
\bar{p} = A \bigg[\dfrac{\bar{\Lambda}_2}{-\bar{\lambda}K + \sqrt{(\bar{\lambda} K + \bar{\Lambda}_2)^2 - 2\bar{\lambda} \bar{\Lambda}_1 \bar{\Lambda}_2 Et}} \bigg]^3 \label{eq:BS_p} \\
\bar{c}_t = \frac{E\lambda(\Lambda_1-1)\bar{c}}{\bar{h}\sqrt{(\lambda K + \Lambda_2)^2 - 2\lambda\Lambda1\Lambda2} Et}
\label{eq:BS_c_t}
\end{gather}

\noindent
where $\bar{\Lambda}_1 = \bar{c} + (1- \bar{c})\alpha\beta^{3/2}\Lambda$ and
$\bar{\Lambda}_2 = \bar{c} + (1- \bar{c})\alpha\beta^{3/2}\Lambda^2$.

\section{Neutral Curves}
Here we present the expressions of the neutral curves for the monotonic and oscillatory case. For the monotonic case the neutral curve is given by,
\begin{equation}
    M_T = \dfrac{\mu_b Pr (\lambda_b K + \Lambda_{2b}h_b)2}{2h_b^2\lambda_b K}
    \left[\dfrac{M}{N}\right]\label{eq:neutral_mono}
\end{equation}

\noindent
where,
\begin{gather}
    M = -[-2E + E^{HH} + E^{CC} + (A^{HH} - D^{CC})k^2 \nonumber \\
    - S^{HH}k^4]^2 
    + [-E^{HH} + E^{CC} - (A^{HH} + D^{CC})k^2 \nonumber \\
    + S^{HH}k^4]^2 
    + 4 E^{CH} (E^{HC} - M_c^{HC}k^2)
\end{gather}
\begin{gather}
    N = \gamma_b \Lambda_{2b} (-E + E^{CC} - D^{CC}k^2)k^2 \nonumber \\
    - E^{CH} ((1-\gamma_r)(\lambda_b K + \Lambda_{2b}h_b) - \gamma_b h_b (1-\alpha\beta^{3/2}\Lambda^2))k^2
\end{gather}

For the oscillatory case the neutral curve is given by,
\begin{gather}
    M_T = \dfrac{2\mu_b Pr (\lambda_b K + \Lambda_{2b}h_b)^2}{h_b^2 \lambda_b K \gamma_b \Lambda_{2b}} \bigg[\dfrac{2E}{k^2} -\dfrac{E^{HH}}{k^2} - \dfrac{E^{CC}}{k^2} \nonumber \\
    - A^{HH} + D^{CC} + S^{HH}k^2 \bigg]\label{eq:neutral_osc}
\end{gather}

\noindent
and the imaginary part of the eigenvalue where the real part is equal to $E$ is given by,
\begin{gather}
    r_I = 2 \big[ -(-E + E^{CC} - D^{CC}k^2)^2 \nonumber \\
    - E^{CH}(E^{HC} + (M_T^{HC*} - M_c^{HC})k^2 \big]^{1/2}
\end{gather}

\noindent
where,
\begin{gather}
    M_T^{CH*} = \dfrac{1}{\gamma_b \Lambda_{2b}} \bigg[\dfrac{2E}{k^2} -\dfrac{E^{HH}}{k^2} - \dfrac{E^{CC}}{k^2} - A^{HH} + D^{CC} \nonumber \\
    + S^{HH}k^2 \bigg] ((1-\gamma_r)(\lambda_b K + \Lambda_{2b}h_b) - \gamma_b h_b (1-\alpha\beta^{3/2}\Lambda^2))
\end{gather}

\bibliography{references}

\end{document}